\documentclass[letterpaper, number, sort&compress]{elsarticle}
\usepackage[utf8]{inputenc}

\usepackage{amsmath}
\usepackage{amsfonts}
\usepackage{amssymb}
\usepackage{booktabs}
\usepackage{bm}
\usepackage[hidelinks]{hyperref}
\usepackage{natbib}
\usepackage{graphicx}
\usepackage{tabularx}
\usepackage{subcaption}
\usepackage[margin=1in]{geometry}
\usepackage{setspace}

\newcommand{\phSc}[2] {^\mathrm{#1}_\mathrm{#2}}

\begin{document}

\begin{frontmatter}

\title{Mesoscale simulation of woven composite design decisions}

\author{Lincoln N. Collins}
\ead{lcolli@sandia.gov}

\author{Scott A. Roberts\corref{sar}}
\ead{sarober@sandia.gov}
\cortext[sar]{Corresponding author}

\address{Engineering Sciences Center, Sandia National Laboratories, Albuquerque, NM, USA}


\begin{abstract}
	Characterizing the connection between material design decisions/parameters and their effective properties allows for accelerated materials development and optimization. We present a global sensitivity analysis of woven composite thermophysical properties, including density, volume fraction, thermal conductivity, specific heat, moduli, permeability, and tortuosity, predicted using mesoscale finite element simulations.  The mesoscale simulations use microscale approximations for the tow and matrix phases. We performed Latin hypercube sampling of viable input parameter ranges, and the resulting effective property distributions are analyzed using a surrogate model to determine the correlations between material parameters and responses, interactions between properties, and finally Sobol' indices and sensitivities. We demonstrate that both constituent physical properties and the mesoscale geometry strongly influence the composite material properties. 
\end{abstract}	

\begin{keyword}
	Polymer-matrix composites (PMCs) \sep Anisotropy \sep Mechanical properties \sep Physical properties \sep Thermal properties \sep Finite element analysis (FEA) 
\end{keyword}

\end{frontmatter}

\section{Introduction}

The extreme environments produced during atmospheric reentry of spacecraft create intense shear forces and aerothermal heating that lead to degradation of exposed surfaces. Thermal protection systems (TPS) are designed to mitigate damage and increase the survivability of these surfaces. Determining the performance of materials in a TPS is a difficult, yet essential, part of vehicle development, where simulation can be used to augment expensive testing. Traditional modeling approaches rely on volume-averaged bulk properties despite minimal understanding of the connection between composite structure and macroscopic properties \cite{Rivier2019,Zhu2019}. Considering the wide range of available constituent materials and composite geometries, a unified examination of the complete set of relevant physical properties has not yet been presented. 

Woven carbon-based phenolic composites are particularly effective at withstanding atmospheric reentry conditions. Carbon fiber filaments are bundled together into yarns, or tows, then woven into a fabric. These fabric sheets are impregnated with a resin matrix, layered, and cured. The matrix phase is comprised of phenolic resin, carbon based filler, and a certain amount of porosity due to manufacturing constraints. Both carbon fibers and the resin matrix exhibit a wide range of properties dependent on precursors and processing \cite{minus2005,Edie1998, Zhang2000}. The use of composites in these systems presents a multi-scale engineering problem; behavior at the microscale drives behavior at the macroscale.

Simulations of an idealized mesostructure can be used to calculate the effective macroscale behavior of the composite. Various approaches exist including rigorous numerical homogenization and two-scale asymptotics \cite{dasgupta1996,carvelli2001}, effective medium theories \cite{hashin1972}, and complete multi-scale simulations \cite{bostanabad2018b,thapa2019}. Rather than coupling with simulation at the microscale, e.g. \cite{Tao2020a}, effective medium theory can approximate behavior of the microscale to inform material properties at the mesoscale \cite{chamis1983} while remaining computationally efficient.

Simple analytical forms describing yarn cross-section and path in a woven fabric are an efficient method of approximating the mesoscale geometry of a composite \cite{naik1994}. Various studies have developed complicated simulated geometries that more closely resemble experimental composites \cite{long2011,rinaldi2012}. Another approach relies on advanced imaging techniques to perform simulations using realistic, imperfect, geometries \cite{drach2014,Tao2020a}. However, addressing the characteristics of a fabric weave arising from manufacturing with a minimal selection of geometric parameters---such as thickness, tow width, waviness, and gap---allows for more concise connections between composite behavior and fabric geometry.

\begin{figure} 
\centering
\includegraphics[width=.75\linewidth]{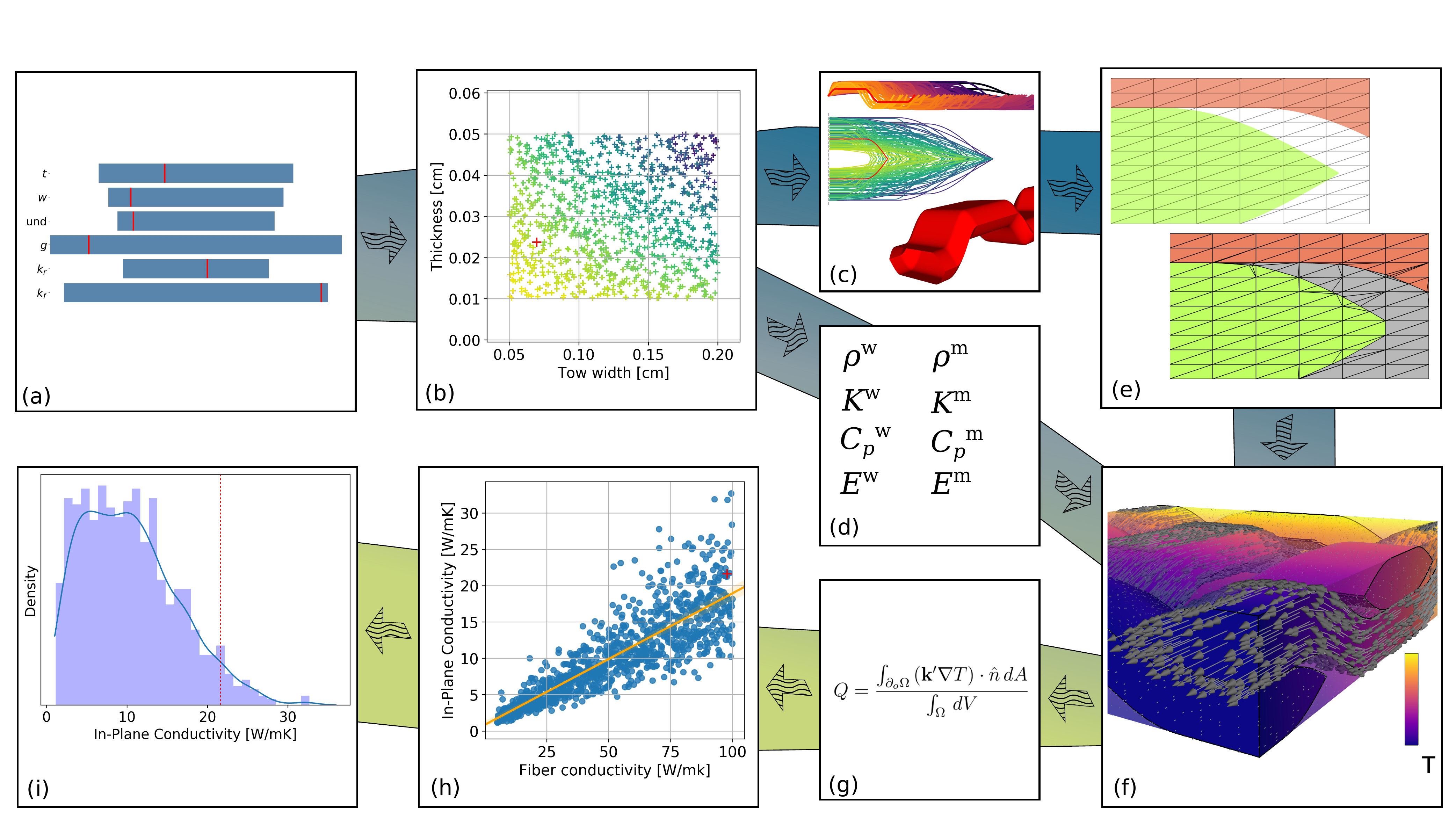}
\caption{Workflow for study exemplified by the calculation of in-plane thermal conductivity. (a) Prescription of uniform ranges for input parameters. (b) Latin hypercube sampling of input space. (c) Generation of yarn paths, cross-sections, and resulting STL geometry for the weave. (d) Up-scaling of microscopic parameters to mesoscale properties of the weave and matrix. (e) Generation of conforming finite element mesh using CDFEM. (f) Example simulation of in-plane heat transport. (g) Calculation of effective property. (h) Polynomial chaos expansion and generation of surrogate model. (i) Associated QoI distribution. Example trials are highlighted in red.}
\label{fig:workflow}
\end{figure}

This study comprehensively explores the impact of design decisions on quantities of interest (QoIs) for TPS performance. The framework for numerically exploring the input parameter space characterizing the composite is presented in \autoref{fig:workflow}. We compute effective properties such as density, thermal conductivity, tortuosity, permeability, elastic moduli, and thermal expansivity. We use Latin Hypercube Sampling (LHS) to explore a design space encompassing constituent material choice, weave geometry, and inclusions for a plain-woven carbon phenolic. The resulting distribution of effective material properties is then analyzed to reveal statistics about the composite design space.

\section{Approach}
An obvious benefit to using analytical, idealized, fabric descriptions is the relative ease in obtaining finite element meshes. Starting with the workflow presented in \autoref{fig:workflow}, a set of geometric parameters is chosen to represent the surfaces of the yarns analytically. A completely closed triangularly faceted surface is described using points sampled uniformly from this surface and then exported as a standard tessellation language (STL) file. Each yarn in the unit cell is described by a separate STL file. 

The set of STL files is used to generate an interface-conformal tetrahedral finite element mesh using the Conformal Decomposition Finite Element Method (CDFEM) \cite{noble2010,roberts2018}. A rectangular domain with dimensions of the unit cell is discretized into tetrahedral elements and acts as the background mesh. The STL files are used to calculate a signed level-set distance function, $\phi$, where $\phi=0$ represents the location of the STL surface. New nodes are added on edges of the background elements where $\phi=0$. Elements containing these new nodes are additionally decomposed into child elements that conform to the interface. Each new element is located in a single material phase. As a result, each feature is produced as a separate block of elements, allowing the desired boundary conditions, interface conditions, and discontinuous material properties to be assigned. Results of this process are presented in \autoref{fig:workflow}.

A local material coordinate frame aligned with the fiber orientation is prescribed on a per-element basis in the weave block. For each element in the yarn block, the centroid position is calculated to obtain the necessary normal and tangential vectors using the derivatives of \autoref{eq:cross_sec_geo} and \autoref{eq:path_geo}. 

Latin Hypercube Sampling (LHS) is implemented to explore the parameter space described by \autoref{tbl:params} and informs a suite of finite element simulations calculating the 19 QoIs presented in \autoref{tbl:props}. A polynomial chaos expansion (PCE) surrogate model is developed, where multivariate orthogonal polynomials are used to describe the output distributions.

The end result is tabulated data representing the results of each simulation, their correlations and sensitivities, as well as the Sobol' indices describing the relations between input and output variations obtained through the PCE regression. Surrogate-based analysis allows for the total variation in a QoI to be distributed among the highest contributing input parameters associated with the calculation. Thus, efforts in optimization, manufacturing, design of experiments, and investigations in minimizing uncertainties in the composite can be focused on material aspects identified by the Sobol index-based sensitivity analysis.

\subsection{Material geometry}
\begin{figure}
	\centering
	\includegraphics[width=0.5\textwidth, draft=false]{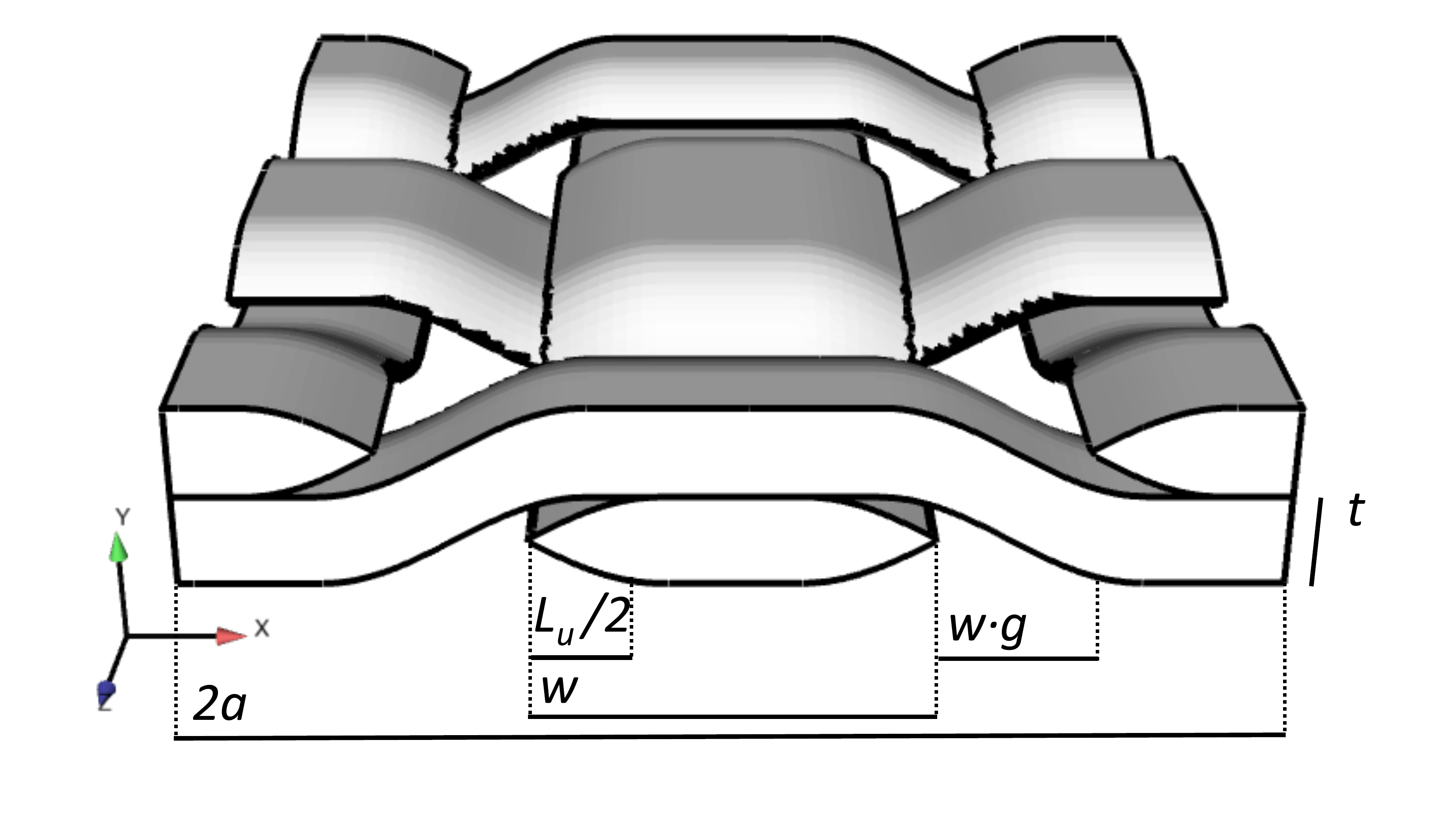}
    \caption{Typical unit cell geometry generated using (\autoref{eq:cross_sec_geo}) and (\autoref{eq:path_geo}) adapted from \cite{naik1994}.}
	\label{fig:nomgeo}
\end{figure}

The unit cell $\Omega$ is comprised of resin-impregnated tows forming the weave $\Omega\phSc{w}{}$ and the resin-based matrix phase $\Omega\phSc{m}{}$. The unit cell maintains in-plane symmetries. A system of piece-wise linear and sinusoidal functions describes the yarn cross-section and centerline path. The functional form of the description is adapted from \cite{naik1994}, and a representative geometry is visualized in \autoref{fig:nomgeo}.

The weave is characterized by tow width $w$ and thickness $t$ using the same dimensions for the warp and weft directions. The undulation $u$ is a shape parameter of the cross section and path, characterized by a length $L_u=u\cdot w$. The gap $g$ describes the space between neighboring tows. Both $u$ and $g$ are expressed by percentage the tow width dimension. For this model, $x$ and $z$ are the two in-plane coordinates and $y$ is normal to the fabric. The yarn cross-section and centerline are defined by
\begin{align}
\label{eq:cross_sec_geo}
f_\text{cross}(x) = \left\{
\begin{array}{lcrcl}
\pm \frac{t}{2}\sin\left(\frac{\pi}{L_u}x\right) & : & 0 \leq & x & \leq \frac{L_u}{2}\\
\pm \frac{t}{2} & : & \frac{L_u}{2} <  & x & \leq \left(w - \frac{L_u}{2}\right) \\
\mp \frac{t}{2}\sin\left(\frac{\pi}{L_u}(x-w)\right) & : & \left(w - \frac{L_u}{2}\right) < & x & \leq w\\
\end{array}
\right.
\end{align}
and
\begin{align}
\label{eq:path_geo}
f_\text{path}(z \pm 2a) = \left\{
\begin{array}{lcrcl}
\frac{t}{2}\sin\left(\frac{\pi}{L_u}z\right) & : & 0 \leq & z & \leq \frac{L_u}{2}\\
\frac{t}{2} & : & \frac{L_u}{2} < & z & \leq a - \frac{L_u}{2} \\
-\frac{t}{2}\sin\left(\frac{\pi}{L_u}(z-a)\right) & : & a - \frac{L_u}{2} < & z & \leq a + \frac{L_u}{2}\\
-\frac{t}{2} & : & a + \frac{L_u}{2} < & z & \leq 2a - \frac{L_u}{2}\\
\frac{t}{2}\sin\left(\frac{\pi}{L_u}(z-2a)\right) & : & 2a - \frac{L_u}{2} & z & \leq 2a
\end{array}
\right..
\end{align}
Here, $a$ represents half of the unit cell width: $a=w(1+g)$. A complete tow surface is constructed by summing the centerline and cross-section functions for $(x,z)$:
\begin{align}
y = f_\text{tow}(x,z) = f_\text{path}(z) \pm f_\text{cross}(x), (x,z) \in (0,w)\times(0,2a).
\end{align}
Waviness $\omega$ describes the average slope of the undulating portion of the yarn in terms of the geometric parameters and will be used for analysis of the composite properties:
\begin{equation}
\label{eq:waviness}
\omega = \frac{2t}{wu}.
\end{equation}

Effective quantities for the composite, matrix, and weave are specified using superscripted $\mathrm{*},\mathrm{m},$ and $\mathrm{w}$ respectively. Where applicable, individual phases (fiber, resin, etc.) are denoted using subscripts followed by the necessary orientation of anistropic properties (axial $\mathrm{a}$, transverse $\mathrm{t}$). The directional composite properties are specified with $\mathrm{ip}$ and $\mathrm{oop}$ for in-plane and out-of-plane directions. For example, the axial thermal conductivity of the fibers is denoted $k\phSc{}{f,a}$, whereas the transverse conductivity of the yarns is denoted with $k\phSc{w}{t}$.

The fiber volume fraction in a unit cell $v\phSc{*}{f}$ is easily calculated by combining the cross-sectional area
\begin{align}
\label{eq:A_cross_section}
A\phSc{w}{}=wt\left(1-u\left(1-\frac{2}{\pi}\right)\right),
\end{align}
with the unit cell dimensions and the fiber packing efficiency $v\phSc{w}{f}$,
\begin{align}
\label{eq:v_frac_a}
v\phSc{*}{f} = v\phSc{w}{f} \frac{\left(1-u\left(1-\frac{2}{\pi}\right)\right)} {(1+g)}.
\end{align}

\subsection{Physical models}
To obtain the quantities of interest (QoIs) describing the composite's thermal, mechanical, and fluid-flow behaviors, distinct finite element simulations are performed using the generated geometry. In this section, we first present the approximations used to derive the mesoscale material properties of the fabric weave and matrix phase from the microscale. Then the governing equations, boundary conditions, and calculations used for obtaining macroscale properties from mesoscale simulations are presented for each set of composite properties.

\subsubsection{Microscale models}
\label{sec:microscale}
The mesoscale simulations consisting of the matrix and weave use closed forms to describe the microscale behavior. The matrix phase is approximated as an ideal multi-phase material using effective media theory. Similarly, the weave phase is approximated as unidirectional arrays of cylinders surrounded by the matrix with locally varying orientation according to the fiber orientation $\frac{\partial f_{\text{path}}}{\partial z}$.

\subsubsection*{Matrix phase}
The matrix phase is treated as an isotropic solid with a uniform distribution of spherical inclusions describing the filler particles and voids. Three phases $i$, resin, filler, and air-filled void, with volume fractions $v\phSc{m}{\mathit{i}}$ are used to calculate the effective matrix properties. The Bruggeman relation \cite{markel2016} is solved numerically to approximate the matrix thermal conductivity, $k\phSc{m}{}$:
\begin{align}
\label{eq:bruggeman_therm}
\sum_{i=1}^{3} v\phSc{m}{\mathit{i}}\frac{k_i-k\phSc{m}{}}{k_i+2k\phSc{m}{}} = 0.
\end{align}

The elastic response is calculated following Berryman \cite{berryman1980b} by using the bulk ($K_i$) and shear ($\mu_i$) moduli of each phase:
\begin{align}\label{eq:berryman}
\sum\limits_{i=1}^N v\phSc{m}{\mathit{i}}(K_i - K\phSc{m}{})P\phSc{*}{\mathit{i}}= 0,\\
\text{and}\quad
\sum\limits_{i=1}^N v\phSc{m}{\mathit{i}}(\mu_i - \mu\phSc{m}{})Q\phSc{*}{\mathit{i}}= 0,
\end{align}
where the shape factors $P\phSc{m}{\mathit{i}}$ and $Q\phSc{m}{\mathit{i}}$ take the form
\begin{align}
P\phSc{m}{\mathit{i}} = \frac{K\phSc{m}{}+\frac{4}{3}\mu\phSc{m}{}}{K_i+\frac{4}{3}\mu\phSc{m}{}},\quad Q\phSc{m}{\mathit{i}} = \frac{\mu\phSc{m}{}+F\phSc{m}{}}{\mu_i+F\phSc{m}{}},
\end{align}
with $F = (\mu\phSc{m}{}/6)\left((9K\phSc{m}{} + 8\mu\phSc{m}{})/(K\phSc{m}{}+2\mu\phSc{m}{})\right)$.

The result of Budiansky \cite{budiansky1970} is used for the effective linear thermal expansion coefficient $\alpha\phSc{m}{}$,
\begin{align}
\label{eq:budiansky}
\alpha\phSc{m}{} = \sum_{i=1}^N  v\phSc{m}{\mathit{i}}(K_i/K\phSc{m}{})(\alpha_i)\left(1-a\phSc{m}{}+a\phSc{m}{}(K_i/K\phSc{m}{})\right)^{-1},
\end{align}
where $K\phSc{m}{}$ and $a\phSc{m}{}=\frac{(1+\nu\phSc{m}{})}{3(1-\nu\phSc{m}{})}$ are computed using \autoref{eq:berryman}. 

\subsubsection*{Yarn phase}
\label{sec:yarn_props}
Assuming that locally the bundles of filaments in the weave behave as hexagonal arrays of unidirectional fibers (phase $\mathrm{f}$) with a prescribed volume fraction $v\phSc{w}{f}$, surrounded by the matrix phase ($\mathrm{m}$), most of the necessary material properties of the tow can be approximated using the approach of Chamis \cite{chamis1983}. The axial $k\phSc{w}{a}$ and transverse $k\phSc{w}{t}$ conductivities are calculated using weighted volume averages:
\begin{align}
\label{eq:voigt_tot}
k\phSc{w}{a} &= k\phSc{}{f,a} v\phSc{w}{f} + k\phSc{m}{} (1-v\phSc{w}{f}),
\end{align}
\begin{align}
\label{eq:chamis_tcond2}
k\phSc{w}{t} &= \left(1-\sqrt{v\phSc{w}{f}}\right)k\phSc{m}{}+\frac{k\phSc{m}{}\sqrt{v\phSc{w}{f}}}{1-\sqrt{v\phSc{w}{f}}\left(1-k\phSc{m}{}/k\phSc{}{f,t}\right)}.
\end{align}
The mechanical properties of the yarn are also weighted volume averages with similar functional forms that consider the anisotropic behavior of the fibers as well:\newline
\begin{tabular}{lr}
Axial Young's Modulus & $E\phSc{w}{a} = E\phSc{}{f,a} v\phSc{w}{f} + E\phSc{m}{} (1- v\phSc{w}{f})$\\
Transverse Young's Modulus & $E\phSc{w}{t} = E\phSc{m}{}\left(1 - \sqrt{v\phSc{w}{f}}(1-E\phSc{m}{}/E\phSc{}{f,t})\right)^{-1}$\\
Axial Shear Modulus & $G\phSc{w}{a}  = G\phSc{m}{}\left(1 - \sqrt{v\phSc{w}{f}}(1-G\phSc{m}{}/G\phSc{}{f,at})\right)^{-1}$ \\
Transverse Shear Modulus & $G\phSc{w}{t} = G\phSc{m}{}\left(1 - \sqrt{v\phSc{w}{f}}(1-G\phSc{m}{}/G\phSc{}{f,tt})\right)^{-1}$  \\
Poisson's Ratio & $\nu\phSc{w}{,at} = \nu\phSc{}{f,at} v\phSc{w}{f} + \nu\phSc{m}{} (1- v\phSc{w}{f})$\\
Axial Thermal Expansivity & $\alpha\phSc{w}{a} = \left(v\phSc{w}{f}\alpha\phSc{}{f,a}E\phSc{}{f,a}+(1-v\phSc{w}{f})\alpha\phSc{m}{}E\phSc{m}{}\right)/E_{y,a}$ \\
Transverse Thermal Expansivity & $\alpha\phSc{w}{t} = \sqrt{v\phSc{w}{f}}+(1-\sqrt{v\phSc{w}{f}})\left(1+v\phSc{w}{f}\nu\phSc{m}{}E\phSc{}{f,a}/E\phSc{w}{a}\right)\alpha\phSc{m}{}$ \\
\end{tabular}

\subsubsection{Mesoscale models and effective macroscale properties}
Upscaled microscale material properties are assigned to the different phases of the mesoscale geometry and used in finite element simulations to determine the effective properties of the composite. Although the equations and calculations are presented for behavior in the $x$-direction, the same calculations are similarly applied to the other directions of the material.

\subsubsection*{Geometric quantities}
Geometric quantities, such as total volume fraction of fibers $v\phSc{*}{f}$, specific surface area of the weave $S\phSc{w}{}$, and specific contact area between yarns $CW\phSc{w}{}$, are calculated using the finite element discretization. Fiber volume fraction is calculated using the volume of the weave phase multiplied by the fiber packing ratio $v\phSc{w}{f}$. The specific surface area denotes the combined surface area of the weave occupying a unit cell. Finally, contact area is calculated by integrating the shared interfaces between the yarns.

Total density of the composite $\rho\phSc{*}{}$ is a consequence of the prescribed geometry and is calculated using a weighted average of the phase densities $\rho\phSc{}{\mathit{i}}$ with their associated volume fractions $v\phSc{*}{\mathit{i}}$: 
\begin{align}
\rho\phSc{*}{}= \sum_i \rho\phSc{}{\mathit{i}} v\phSc{*}{\mathit{i}}.
\end{align}

\subsubsection*{Thermal properties}
Thermal conductivity is calculated in the in-plane $\mathrm{ip}$ and out-of-plane $\mathrm{oop}$ directions by solving
\begin{align}
\nabla\cdot\left(\mathbf{k}' \nabla T\right) = 0\text{    in   }\Omega',
\end{align}
where $\mathbf{k}'=\mathbf{k}\phSc{w}{},\mathbf{k}\phSc{m}{}$ denotes the thermal conductivities of the weave ($\Omega\phSc{w}{}$) and matrix ($\Omega\phSc{m}{}$) and $T$ is the temperature. In each direction, a thermal gradient is applied by enforcing Dirichlet boundary conditions on opposing faces of the domain. Symmetry is enforced via adiabatic boundary conditions on the remainder of the unit cell boundary. The resulting mean heat flux $Q$ across the low-temperature boundary $\partial_o\Omega$ with outward normal $\hat{n}$,
\begin{align}
Q = \frac{-\int_{\partial_o\Omega} \left(\mathbf{k}' \nabla T\right)\cdot\hat{n} \,dA}{\int_{\partial_o\Omega}\,dA},
\end{align}
allows for the calculation of effective in-plane thermal conductivity $k\phSc{*}{ip}$:
\begin{align}
\label{eq:eff_cond}
k\phSc{*}{ip} = -\frac{L\phSc{}{ip}}{\Delta T}Q,
\end{align}
where $L\phSc{}{ip}=2a=2w(1+g)$ denotes the in-plane unit cell dimension. In the out-of-plane direction, the same equations and boundary conditions are applied with suitable alterations, where $L\phSc{}{oop}=2t$.

Specific heat $C\phSc{*}{}$ is calculated using the mass fraction and specific heat, $C\phSc{}{\mathit{i}}$, for each constituent:
\begin{align}
C\phSc{*}{} = \frac{1}{\rho\phSc{*}{}}\sum_i C_i \rho\phSc{}{\mathit{i}} v\phSc{*}{\mathit{i}}.
\end{align}

\subsubsection*{Mechanical properties}
Effective mechanical behavior of the composite is obtained from simulations carried out under the assumptions of linear elasticity and that the composite is transversely isotropic. Thus, five independent mechanical parameters (in-plane and out-of-plane Young's moduli, the $xy$ and $xz$ shear moduli, and $yx$ Poisson's ratio) and two values of thermal expansivity (in-plane and out-of-plane) are necessary to fully characterize the composite. To obtain the relevant Young's moduli and associated Poisson's ratios, the unit cell is subjected to prescribed displacements in-plane and out-of-plane separately. In all simulations, linear elastic equilibrium is found by solving for the resulting Cauchy stress, $\bm{\sigma}$, satisfying:
\begin{align}
\label{eq:lin_elast}
\nabla\cdot\bm{\sigma} = 0,
\end{align}
where $\bm{\sigma}$ is related to the strain tensor $\bm{\varepsilon} = \frac{1}{2}\left(\nabla \bm{u} + (\nabla \bm{u})^\intercal\right)$ associated with a displacement field $\bm{u}$ through the fourth-order elastic stiffness tensor $\mathbb{C}$ incorporating relevant material constants:
\begin{align}
\bm{\sigma} = \mathbb{C}:\left(\bm{\varepsilon}-\bm{\alpha}\Delta T\right),
\end{align}
where $\bm{\alpha}$ is the coefficient of thermal expansion (CTE) tensor.

Hashin \cite{hashin1972} presents methods for volume averaging effective properties in the absence of periodic boundary conditions and rigorous numerical homogenization (e.g. \cite{lomov2007}). For example, to determine the composite Young's modulus $E\phSc{*}{\mathit{xx}}$ and Poisson's ratios $\nu\phSc{*}{\mathit{xy}},\nu\phSc{*}{\mathit{xz}}$ involved with in-plane loading, a fixed displacement $u_x$ is applied in the load direction across every boundary of the domain $\partial\Omega$, which remain traction-free in the other directions:
\begin{align}
& u_x(\partial\Omega) = \varepsilon^0_{xx}x\quad \left(\bm{\sigma}\cdot\hat{n}\right)_y(\partial\Omega) = \left(\bm{\sigma}\cdot\hat{n}\right)_z(\partial\Omega) = 0.
\end{align}
Thus, the effective properties arising from this loading are calculated with:
\begin{align}
E\phSc{*}{\mathit{xx}} = \frac{\bar{\sigma}_{xx}}{\varepsilon^0_{xx}}\quad \nu\phSc{*}{\mathit{xy}} = - \frac{\bar{\varepsilon}_{yy}}{\varepsilon^0_{xx}} \quad \nu\phSc{*}{\mathit{xz}} = - \frac{\bar{\varepsilon}_{zz}}{\varepsilon^0_{xx}}.
\end{align}
Here, the enforced boundary condition is denoted with $\cdot^0$ and the corresponding volume-averaged quantities are denoted with $\bar{\cdot}$. Volume-averaged quantities are calculated numerically.

The shear modulus, $G_{xz}$, is calculated similarly: 
\begin{align}
& u_x(\partial\Omega) = \varepsilon^0_{xz}z,\quad u_y(\partial\Omega) = 0, \quad u_z(\partial\Omega) = \varepsilon^0_{xz}x.
\end{align}
Here, $\varepsilon^0_{xz}$ represents the applied shear displacement across the surface of the unit cell. The only non-zero stress value is used to calculate the effective shear modulus:
\begin{align}
G\phSc{*}{\mathit{xz}} = \frac{\bar{\sigma}_{xz}}{2\varepsilon^0_{xz}}.
\end{align}
The other shear configuration, $xy=zy$, follows the same formulation with an altered orientation. 

To calculate the effective thermal expansivity, a uniform temperature difference ($\Delta T$) is applied. We assume that CTE is independent of temperature. Effective values of the CTE are calculated with:
\begin{align}
\alpha\phSc{*}{ip} = \frac{\bar{\varepsilon}_{11}}{\Delta T} = \frac{\bar{\varepsilon}_{33}}{\Delta T},\quad \alpha\phSc{*}{oop}=\frac{\bar{\varepsilon}_{22}}{\Delta T}.
\end{align}

\subsubsection*{Permeability}
Permeability of the composite, absent resin, is relevant for both manufacturing and gas transpiration. Steady-state Stokes flow is modeled through the inter-yarn space typically occupied by the matrix phase:
\begin{align}
\nabla \cdot \bm{v} &= 0, \\ 
\nabla\cdot \bm{T} &= 0,
\end{align}
where $\bm{v}$ is the fluid velocity. The stress tensor, $\bm{T} = -p\bm{I} + \mu\left(\nabla\bm{v}+(\nabla\bm{v})^\intercal\right)$, consists of hydrostatic contributions from the pressure $p$ and terms arising from the fluid viscocity $\mu$. We forgo coupling with porous transport through the yarns because of their low permeability compared to the fluid domain and limited influence on the averaged behavior \cite{simacek1996,nedanov2002,zeng2014}.

Boundary conditions prescribe a pressure gradient in the direction of permeability measurement, e.g., in the $x$-direction, $\Delta P$ is applied across $x=[0,L_x]$. We prescribe an open flow boundary condition at the inlet and outlet, corresponding to the high and low pressure boundaries, with the assumption that the flow is well established and perpendicular to the domain boundary, i.e., non-normal components of the velocity are forced to zero. The flux boundary condition $\bm{q}$ applied on the velocity uses prescribed boundary pressure value and fluid viscosity. It accounts for the stress contribution from the inlet/outlet flows in the equilibrium equation while removing the normal component of the viscous stresses (and other non-pressure stresses): 
\begin{align}
\bm{q} = -p_o\bm{n}+\bm{n}\cdot\bm{\tau}\cdot(\bm{I}-\bm{n}\bm{n}),
\end{align}
where $p_o$ is the prescribed pressure, $\bm{n}$ is the surface normal, and $\bm{\tau}$ is the sum of non-pressure stresses.

In the interior, no-slip conditions are used for the yarn surfaces,
\begin{align}
\bm{v}=0\text{ on }\left(\partial\Omega\phSc{m}{}\setminus\partial\Omega\right),
\end{align}
and zero-penetration conditions are used for the remaining boundaries of the fluid domain:
\begin{align}
v_y = 0\text{ on }\partial\Omega\phSc{m}{\mathit{y}},\\
v_z = 0\text{ on }\partial\Omega\phSc{m}{\mathit{z}}.
\end{align}
The calculation of the effective permeability relies on Darcy's law:
\begin{align}
\label{eq:darcy}
\bm{q} = - \frac{\kappa}{\mu}\nabla p,
\end{align}
such that the effective permeability is obtained with:
\begin{align}
\kappa\phSc{w}{ip} = \bar{v}_x \frac{\mu L_x}{\Delta p},
\end{align}
where $\bar{v}_x$ is the superficial velocity at the outlet wall, i.e., the fluid velocity averaged across the entire boundary at the outlet $\partial\Omega_o = \partial\Omega\phSc{m}{\mathit{o}} \cup \partial\Omega\phSc{w}{\mathit{o}}$.

In considering fluid flow through the weave, tortuosity is another metric representing relative mean path length and indicates the ease of transport through the geometry. It is calculated by modeling species transport through the matrix domain $\Omega\phSc{m}{}$ with matrix diffusivity set to one $D\phSc{m}{} = 1$ and diffusivity of the fiber phase set to zero $D\phSc{}{f} = 0$. The same effective media theories used for conductivity approximate the diffusivity of the yarn phase:
\begin{align}
\label{eq:voigt_diff}
D\phSc{w}{a} &= D\phSc{}{f} v\phSc{w}{f} + D\phSc{m}{} (1-v\phSc{w}{f}),
\end{align}
\begin{align}
\label{eq:chamis_diff}
D\phSc{w}{t} &= \left(1-\sqrt{v\phSc{w}{f}}\right)D\phSc{m}{}+\frac{D\phSc{m}{}\sqrt{v\phSc{w}{f}}}{1-\sqrt{v\phSc{w}{f}}\left(1-D\phSc{m}{}/D\phSc{}{f}\right)}.
\end{align}

The tortuosity is calculated using
\begin{align}
\tau = \frac{(1-v\phSc{*}{f})}{D\phSc{*}{}},
\end{align}
where the effective diffusivity of the unit cell follows from a similar calculation to thermal conductivity: 
\begin{align}
D\phSc{*}{ip} = -\frac{L\phSc{}{ip}}{\Delta c}Q_i,
\end{align}
where $c$ now represents an ionic concentration and $Q_i$ is the associated flux. 
\subsubsection{Implementation details}
The global sensitivity analysis is performed using Dakota \cite{dakotausers}. Studies are split into increments of 250, 500, and 1000 samples to indicate the convergence of distributions and sufficient coverage of the inputs. The subsequent PCE performed is second order, using least angle regression and variance based decomposition of the results. 

The finite element mesh resolution is set by discretizing the unit cell length, $2a$, into 140 intervals and enforcing the out-of-plane element edge to be a third of the in-plane edge length. Typical unit cell meshes produced using CDFEM in this study have roughly two million elements. Mesh resulution was deemed sufficient for problem convergence and accurate surface quality from the CDFEM geometry.

The majority of the physical models are solved using a Galerkin finite-element scheme with linear basis functions through the SIERRA/Aria multi-physics module \cite{Aria4.56}. A Newton method is used for the nonlinear iterative solve, and the resulting linear system is solved with using the generalized minimal residual (GMRES) method and a multilevel preconditioner. For the permeability calculations, a more robust ILUT preconditioning method is used for the linear solve and both Streamline-Upwind/Petrov-Galerkin (SUPG) and Pressure-Stabilizing/Petrov-Galerkin (PSPG) are implemented. The mechanical calculations are performed using the SIERRA/Adagio module, implementing its implicit solver utilizing a preconditioned conjugate gradient method \cite{Adagio4.56}. 

\section{Material parameters}
For this study, we focus on a generic plain-woven carbon phenolic composite with a prescribed porosity and amount of carbon-based filler in the matrix. Carbon fiber properties are based off those typical for polyacrylonitrile (PAN)-based fibers. Variation in precursor materials, processing, and manufacturing conditions lead to differences in fiber microstructure and, accordingly, a wide range of fiber properties. Similarly, the phenolic resin is strongly dependent on processing and cure conditions. As such, resin properties are roughly derived from reported values in the literature, although there is a significant shortage of available data. Finally, filler material is approximated using properties spanning various forms of carbon. 

For the global sensitivity analysis, the input parameter ranges describing the geometry and constituent properties are chosen to represent the full range of available materials and thus, the resulting full response range for the composite. In that respect, we pose the parametric study to cover the design space for a chosen woven composite, where the response distributions can be used to tune the composite design for a specific property and application. Accordingly, the input space (\autoref{tbl:params}) is comprised of uniform ranges determined from extremal values found in literature. Associated sources are annotated in \autoref{tbl:params} and span both experimental and simulation studies. Elsewhere, ranges are chosen using engineering estimates of the variation in material choice. The following discussion summarizes the rationale for and general trends of the values chosen.

The analytical geometry model (\autoref{eq:cross_sec_geo},\autoref{eq:path_geo}) requires the tow width $w$, thickness $t$, undulation parameter $u$, and gap $g$. Depending on weave process, filament count, fiber dimensions, and cure pressures, these parameters vary significantly for plain woven carbon fabrics. Ranges for tow dimensions are expansions from those presented by Naik \cite{naik1995,naik1997}. The undulation parameter varies between 0.3 and 1, corresponding to a range of undulation representing sharply crimped fabrics for the low parameter value and uncontrained tows for the high parameter value. The gap parameter has been chosen to represent the range of a tightly woven fabric and a very loose fabric. 

Fiber packing ratio designates the packing of filaments within the yarn, and is chosen to represent a wide range of values up to close-packed. Investigations into non-uniformities of fiber packing have been performed elsewhere, e.g., \cite{faes2016,muhlstadt2017}. Porosity is taken as fraction of matrix volume, and filler loading is a fraction of resin by mass. 

The range of fiber thermal conductivity values is suitably large to represent variations found in literature \cite{minus2005,pradere2009,emmerich2014,villiere2013}. Depending on the processing of the fibers, a large degree of anisotropy can occur in the carbon filaments due to their microstructure. Fiber anisotropy $\gamma\phSc{}{f}$ denotes the ratio between the transverse ($\mathrm{t}$) and axial ($\mathrm{a}$) conductivity in the filaments and is roughly based off of available data for transverse measurements on single fibers. For example, Villiere \cite{villiere2013} measures Toray T300 at 8.8 W/$[\text{m}\cdot\text{K}]$ in the axial direction, but 2.0 W/$[\text{m}\cdot\text{K}]$ in the transverse direction, yielding a fiber anisotropy of 0.23. We consider the range encompassing highly directional and isotropic fibers. Neat phenolic resin properties are limited in availability and the range presented is based off of assumed properties.

By treating both the filler and resin as isotropic, three elastic constants are necessary for each phase: Young's modulus, shear modulus, and Poisson's ratio. Ranges are chosen as approximations of the limited available data. These six property ranges---for Young's modulus, Poisson's ratio, and CTE---are largely estimated for both the phenolic and filler. Under the assumption that the fibers are transversely isotropic, five elastic constants and two thermal expansivities are necessary. The elastic moduli in the axial direction is obtained through experimental studies \cite{liu2012, miyagawa2005,naik1995, hashin1979}. As with conductivity, there is a strong directionality to the elastic behavior and the transverse modulus is the focus of many studies and simulations that provide a basis for the range explored \cite{miyagawa2005, naik1995, hashin1979}. Two separate Poisson's ratios and a shear modulus are used to complete the characterization of the filaments. Typically, carbon sheets are concentrically wrapped in the filaments, so it is common for fibers to initially contract axially upon heating \cite{Trinquecoste1996}. Thus, the axial coefficient of thermal expansivity spans both positive and negative values in our study.

\begin{table}
	\centering
	\renewcommand{\arraystretch}{0.85}
\begin{tabular}{lllrrr}
\toprule
Parameter &       Symbol &Units &          Min &          Max & Ref. \\
\midrule
\textbf{Geometry etc.} & & & & & \\
Tow Width                        & w  &      [cm] &     0.05 &     0.2   &   \cite{naik1995,naik1997}\\
Thickness                        & t  &      [cm] &     0.01 &    0.05   &   \cite{naik1995,naik1997}\\
Undulation                       & u  &           &     0.30 &   1.0   &   \cite{naik1995,naik1997}\\
Tow Gap                          & g   &            &     0.00 &   0.7   &  \\
Fiber Packing Ratio              & $v\phSc{w}{f}$  &            &      0.5 &     0.9 &\\
Porosity                         & $v\phSc{m}{pore}$  &            & 0 &    0.2 &\\
Filler Loading                   & $m\phSc{m}{fill}$  &            & 0 &     0.2 &\\
Resin Density                    & $\rho\phSc{}{res}$  &[g/cm$^3$] &      1.2 &          1.7  & \\
Fiber Density                    & $\rho\phSc{}{f}$  & [g/cm$^3$] &      1.7 &          1.9 & \cite{minus2005}\\
Filler Density                   & $\rho\phSc{}{fill}$  & [g/cm$^3$] &      1.4 &          2.3  & \cite{nistwebbook} \\
\midrule

\textbf{Thermal} & & & & & \\
Resin Specific Heat              & $C\phSc{}{res}$  &    [J/(kg$\cdot$K)] &      1300 &         1700  & \\
Fiber Specific Heat              & $C\phSc{}{f}$  &    [J/(kg$\cdot$K)] &      600 &        800  & \\
Filler Specific Heat             & $C\phSc{}{fill}$  &    [J/(kg$\cdot$K)] &       1300 &         1800  & \cite{nistwebbook} \\
Resin Conductivity               & $k\phSc{}{res}$  &     [W/(m$\cdot$K)] &     0.2 &         0.6  & \\
Fiber Conductivity               & $k\phSc{}{f,a}$  &     [W/(m$\cdot$K)] &       5 &        100  & \cite{minus2005,pradere2009,emmerich2014,villiere2013}\\
Fiber Anisotropy  (conductivity) & $\gamma\phSc{}{f}$  &            &     0.1 &         1  & \cite{villiere2013} \\
Filler Conductivity              & $k\phSc{}{fill}$  &     [W/(m$\cdot$K)] &     0.2 &         100  & \cite{nistwebbook} \\
\midrule

\textbf{Mechanics} & & & & & \\
Resin Young's Modulus             & $E\phSc{}{res}$  &      [GPa] &      2 &        5  & \\
Fiber Young's Modulus (a)         & $E\phSc{}{f,a}$  &      [GPa] &       200 &       600 & \cite{liu2012, miyagawa2005,naik1995, hashin1979}\\
Fiber Young's Modulus (t)         & $E\phSc{}{f,t}$  &      [GPa] &      5 &        50  & \\
Filler Young's Modulus            & $E\phSc{}{fill}$  &      [GPa] &      5 &          50  & \\
Fiber Shear Modulus (at)         & $\mu\phSc{}{f,at}$  &      [GPa] &      3 &           30  & \cite{miyagawa2005, naik1995, hashin1979} \\
Resin Poisson's Ratio             & $\nu\phSc{}{r}$  &            &     0.25 &       0.35 & \\
Fiber Poisson's Ratio (tt)        & $\nu\phSc{}{f,tt}$  &            &     0.25 &        0.5  & \cite{miyagawa2005, naik1995, hashin1979} \\
Fiber Poisson's Ratio (at)        & $\nu\phSc{}{f,at}$  &            &     0.25 &     0.35  & \cite{miyagawa2005, naik1995, hashin1979} \\
Filler Poisson's Ratio            & $\nu\phSc{}{fill}$  &            &     0.25 &      0.35 & \\
Resin CTE                        & $\alpha\phSc{}{res}$  &      [$\times 10^{-6}$ K$^{-1}$]      &   50 &   100 & \\
Fiber CTE (a)                    & $\alpha\phSc{}{f,a}$  &      [$\times 10^{-6}$ K$^{-1}$]        &  -0.1 &  0.1& \cite{pradere2008} \\
Fiber CTE (t)                    & $\alpha\phSc{}{f,t}$  &      [$\times 10^{-6}$ K$^{-1}$]        &   5 &    10 & \cite{pradere2008} \\
Filler CTE                       & $\alpha\phSc{}{fill}$  &      [$\times 10^{-6}$ K$^{-1}$]        &   1 &  10  & \\
\bottomrule
\end{tabular}\\
\caption{Summary of input parameter ranges for GSA. Note that engineering estimates are used where there is no reference listed.}
\label{tbl:params}
\end{table}

\section{Global Sensitivity Results}
	\begin{table}
		\centering
		\begin{tabular}{llllll}
			\toprule
			Response                     & Symbol &      Units &     Mean &     CV &      DoN \\
			\midrule
			\textbf{Geometry etc.} & & & & & \\
			Volume Fraction              & $v\phSc{*}{f}$  &           &    0.405 &  0.247 &   0.0214 \\
			Specific Surface Area        & $S\phSc{w}{}$  &            &       56 &  0.445 &    0.212 \\
			Specific Contact Area        & $CA\phSc{w}{}$  &            &     19.7 &  0.525 &    0.303 \\
			Density                      & $\rho\phSc{*}{}$  & [g/cm$^3$] &     1.53 & 0.0681 &  0.00146 \\
			
			\midrule
			
			\textbf{Fluid flow} & & & & & \\
			In-Plane Tortuosity          & $\tau\phSc{*}{ip}$  &            &     1.23 & 0.0552 &  0.00877 \\
			Out-of-Plane Tortuosity      & $\tau\phSc{*}{oop}$  &            &     1.58 &   0.17 &   0.0312 \\
			In-Plane Permeability        & $\kappa\phSc{w}{ip}$  &   [cm$^2$] & 1.68$\times 10^{-5}$ &  0.972 &     0.19 \\
			Out-of-Plane Permeability    & $\kappa\phSc{w}{oop}$  &   [cm$^2$] & 2.41$\times 10^{-5}$ &   1.32 &    0.897 \\
			\midrule
			
			\textbf{Thermal} & & & & & \\
			Specific Heat                & $C\phSc{*}{}$  &    [J/(kg$\cdot$K)] & 1.12$\times 10^{3}$ & 0.0934 &  0.00527 \\
			In-Plane Conductivity        & $k\phSc{*}{ip}$  &     [W/(m$\cdot$K)] &     10.4 &  0.567 &  0.00406 \\
			Out-of-Plane Conductivity    & $k\phSc{*}{oop}$  &     [W/(m$\cdot$K)] &     2.37 &  0.789 &    0.216 \\
			\midrule 
			
			\textbf{Mechanical} & & & & & \\
			In-plane Young's Modulus     & $E\phSc{*}{ip}$  &      [GPa] &     31.3 &  0.509 &   0.0603 \\
			Out-of-plane Young's Modulus & $E\phSc{*}{oop}$  &      [GPa] &     6.73 &  0.396 &   0.0581 \\
			Shear Modulus ($xy$)           & $\mu^*_{xy}$  &      [GPa] &     3.74 &  0.388 &   0.0438 \\
			Shear Modulus ($xz$)           & $\mu^*_{xz}$  &      [GPa] &     2.89 &  0.401 &  0.00963 \\
			Poisson's Ratio ($yx$)         & $\nu^*_{yx}$  &            &    0.158 &  0.451 &    0.118 \\
			Poisson's Ratio ($zy$)         & $\nu^*_{zy}$  &            &     0.47 &  0.151 &   0.0492 \\
			In-plane CTE                 & $\alpha\phSc{*}{ip}$  &   [$\times 10^{-6}$ K$^{-1}$]          &     12.9 &  0.494 & 0.158 \\
			Out-of-Plane CTE             & $\alpha\phSc{*}{oop}$  &    [$\times 10^{-6}$ K$^{-1}$]         &     52.6 &  0.334 & 0.079 \\
			\bottomrule
		\end{tabular}
		\caption{Output ranges for the QoIs examined in the GSA.}
		\label{tbl:props}
	\end{table}

\autoref{tbl:props} summarizes the results of the global sensitivity analysis (GSA). The coefficient of variation (CV) is the standard deviation normalized by the mean and illustrates the normalized spread in results. Large values, such as for out-of-plane permeability, suggest a strong sensitivity with respect to input parameters. The degree of nonlinearity (DoN) measures the deviation between the nominal property value (the effective property calculated using the mean of the input parameters) and the mean of the QoI. It is expected that properties more dependent on constituent material properties, rather than geometry, will be more linear and have a lower DoN. Comparing in-plane and out-of-plane conductivity, the DoN for the out-of plane value is much higher, suggesting the unit cell geometry dominates over material properties.

\begin{figure}
	\centering
	\includegraphics[width=0.4\linewidth]{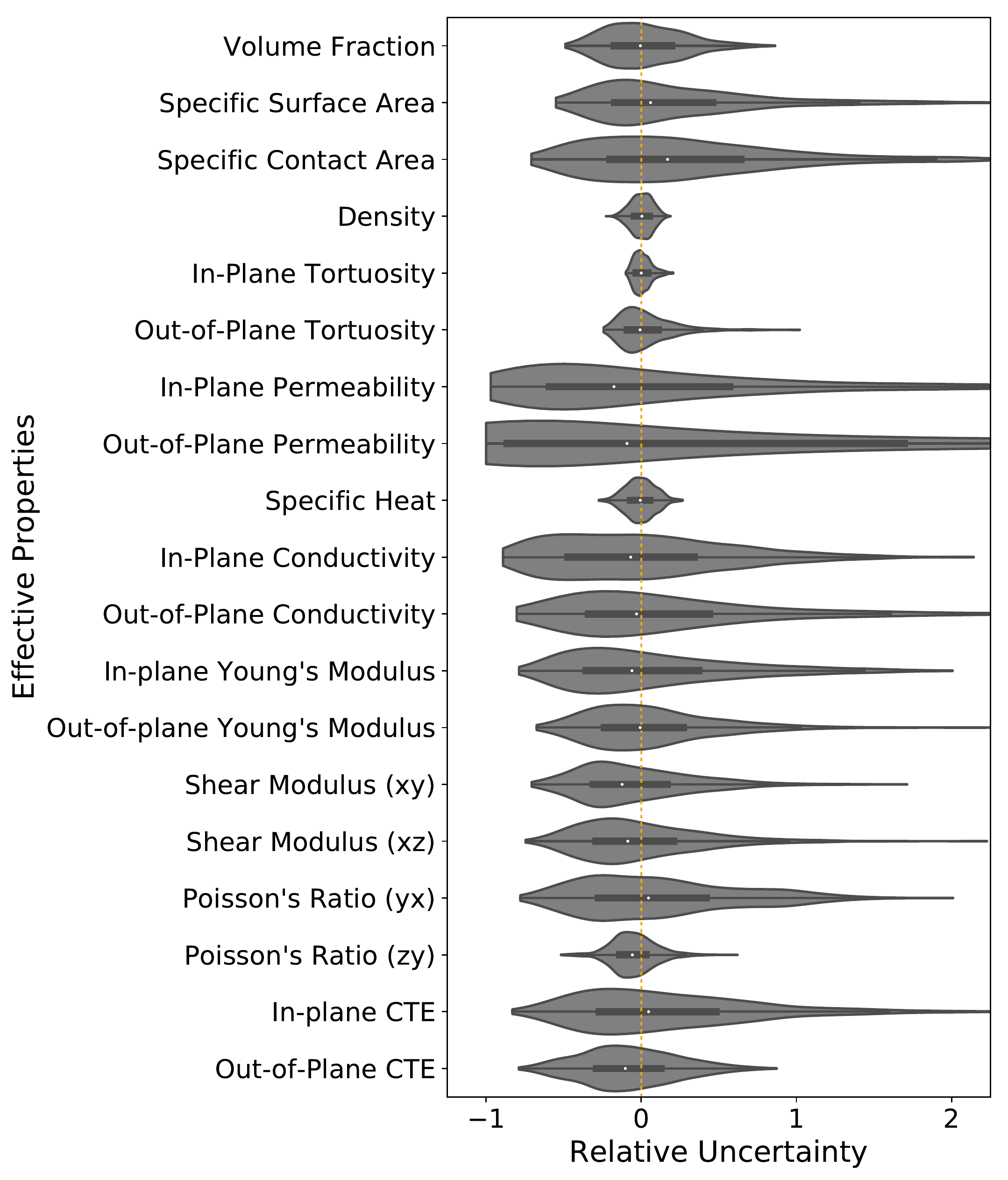}
	\caption{Resulting output distributions from global sensitivity study, normalized to the nominal values. Results falling in the middle two quartiles are captured by boxes containing a point at the mean of the results. The orange dotted line denotes the nominal value at 0.}
	\label{fig:violinplot_total}
\end{figure}

Normalized distributions of the outputs are visualized in \autoref{fig:violinplot_total} and represent how sensitive a QoI is to input variation. The DoN is conveyed through the separation between the mean and nominal markers on the the violin plot. Certain properties only rely on the volume fraction of phases (such as density or specific heat) whereas others are highly dependent on the geometry of the unit cell (such as permeability) and thus are assumed to be more nonlinear. Permeability has the widest distribution because of the high sensitivity to select geometrical parameters, namely the gap between yarns. Conversely, the low normalized standard deviations seen in density and specific heat when compared to the volume fraction distribution is likely due to the low variation in the constituent properties sampled. Although they have similar computations, tortuosity and conductivity show grossly different distributions, where the latter is much wider due to the constituent material properties being sampled, outweighing the role of the geometry. Finally, the difference between in-plane and out-of-plane property distributions indicates the importance of anisotropy in composite modeling. 

\begin{figure}
	\centering
	\begin{subfigure}{0.4\textwidth}
		\centering
		\includegraphics[width=\linewidth]{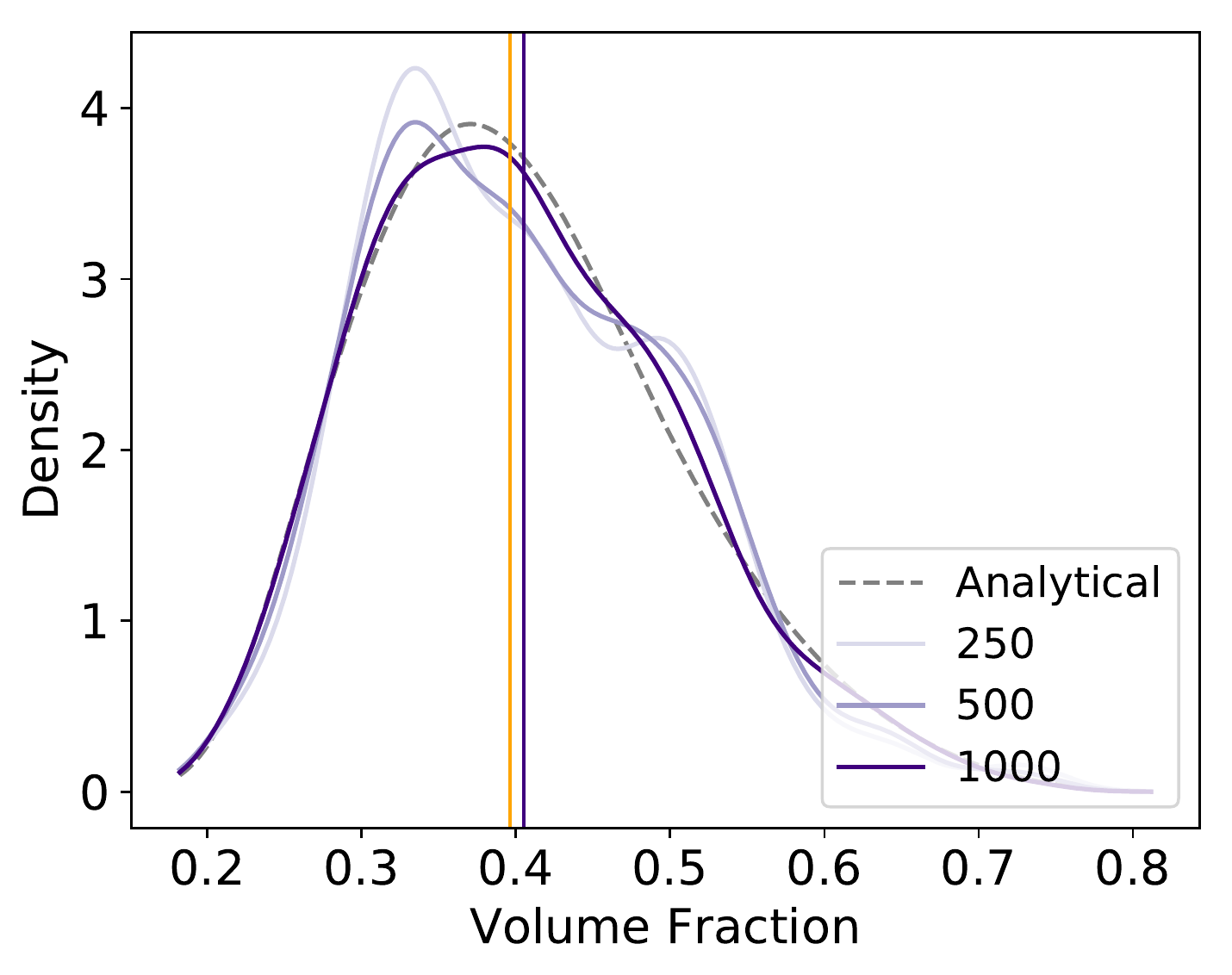}
		\caption{}
		\label{fig:dens_vfrac}
	\end{subfigure}
	\begin{subfigure}{0.4\textwidth}
		\centering
		\includegraphics[width=\linewidth]{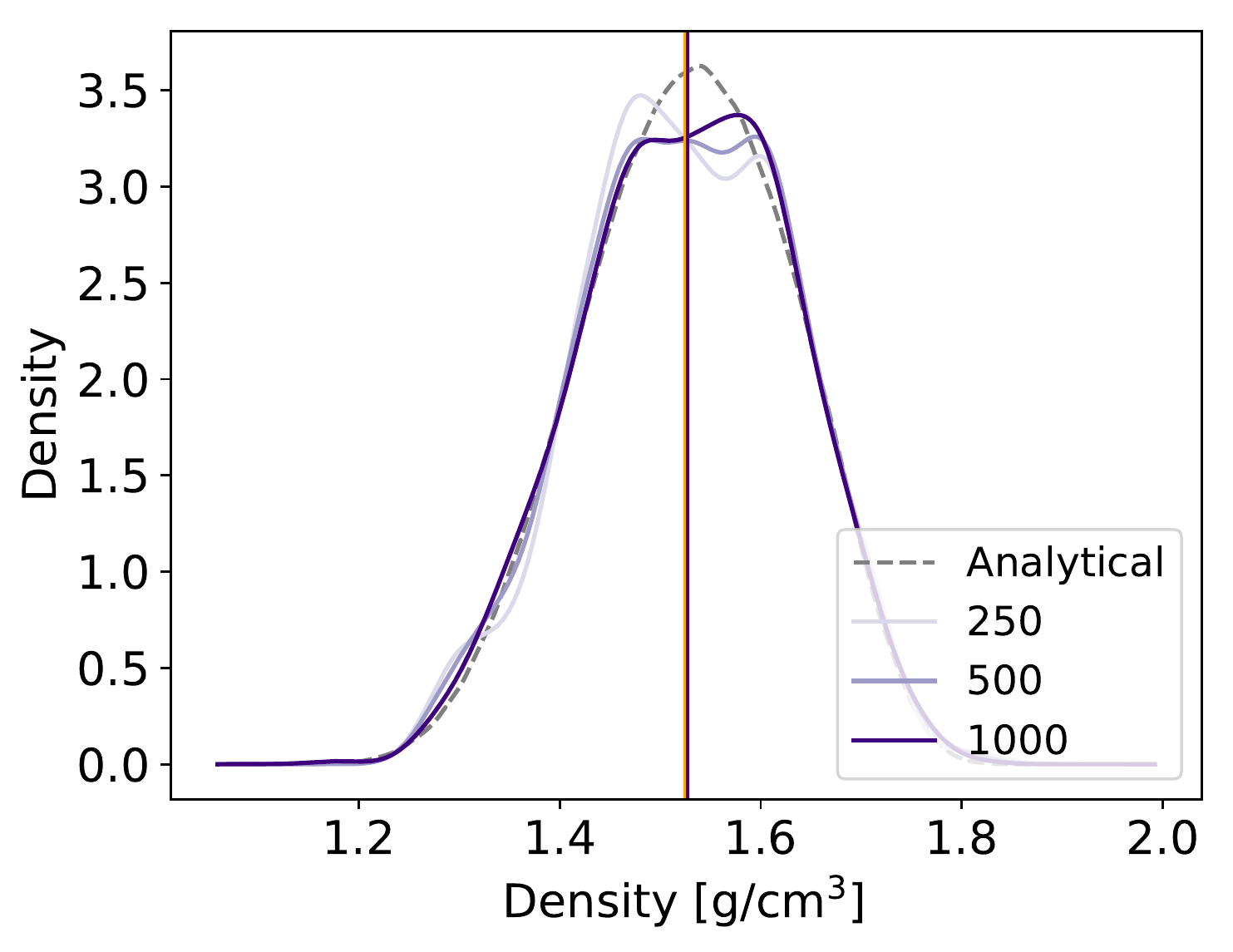}
		\caption{}
		\label{fig:dens_density}
	\end{subfigure}
	\caption{Distributions as a function of LHS size for (\subref{fig:dens_vfrac}) volume fraction and (\subref{fig:dens_density}) density. Vertical lines represent the mean of each distribution, and the orange line refers to the value derived from the nominal set of parameters.}
	\label{fig:dist_comp}
\end{figure}

Select output distributions such as fiber volume fraction and density can be calculated using the analytical description of the unit cell. \autoref{fig:dist_comp} compares the obtained distributions for different sample sizes and the analytical equivalents, illustrating the convergence of the LHS method. The density distribution is narrower and normal than volume fraction from the additional sampling of constituent densities following the central limit theorem.

\begin{figure}
	\centering
	\includegraphics[width=0.8\linewidth]{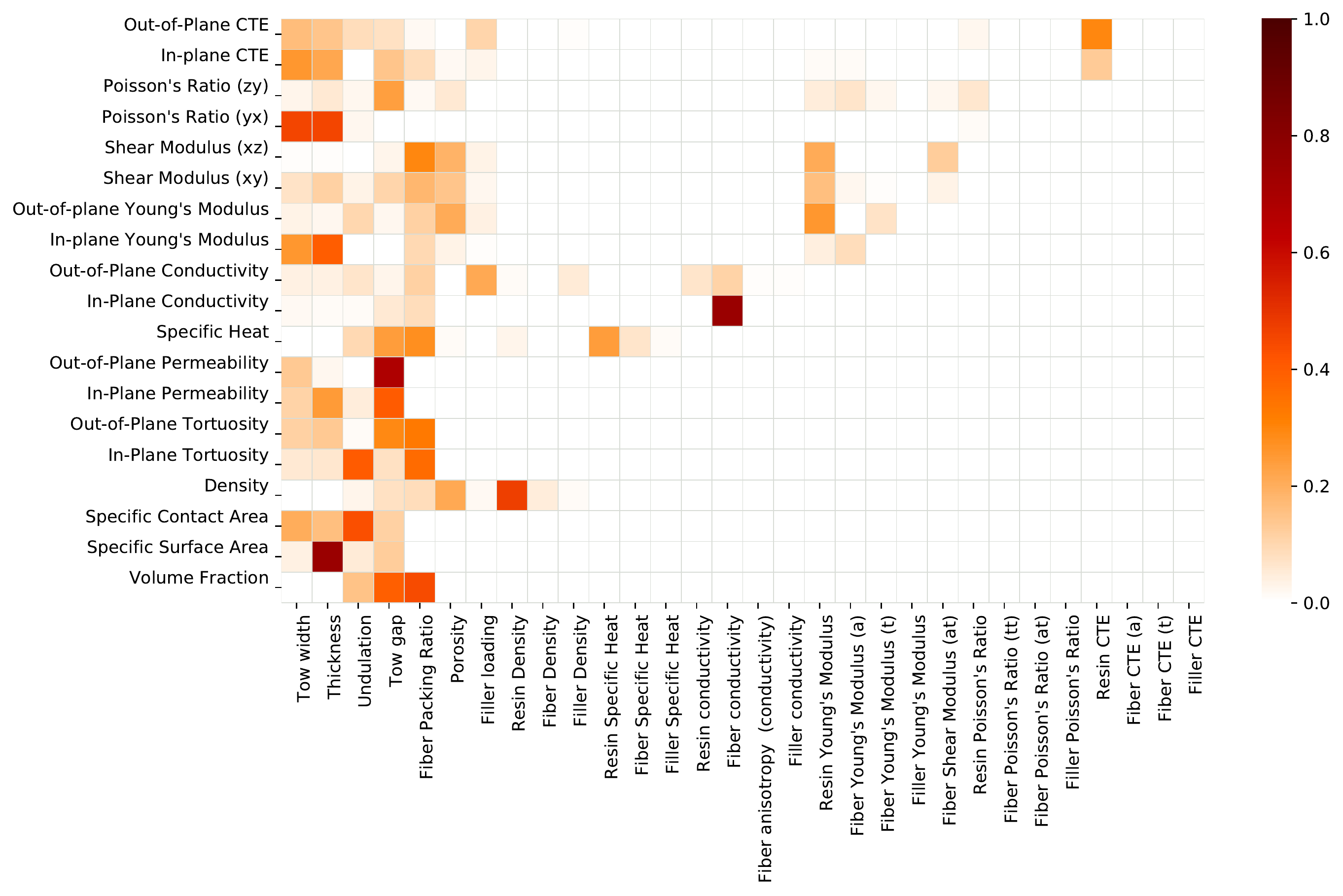}
	\caption{Main effect Sobol' indices obtained using the PCE-developed surrogate model and \autoref{eq:soboldef} \cite{dakotatheory}.}
	\label{fig:heat_sobol}
\end{figure}

Sobol' indices (\autoref{eq:soboldef}) are an essential metric for analyzing large parametric studies by identifying contributions to QoI variations from specific input parameters. For a given output, $Y$, the main effect Sobol' index with respect to input $x_i$ is denoted as the ratio between the variance of the conditional expectation assuming only contribution from $x_i$, $\mathrm{Var}_{x_i}\left[E(Y|x_i)\right]$, and the total variation $\mathrm{Var}(Y)$ \cite{dakotatheory}:
\begin{align}
\label{eq:soboldef}
S_i = \frac{\mathrm{Var}_{x_i}\left[E(Y|x_i)\right]}{\mathrm{Var}(Y)}.
\end{align}
The total variation is normalized across each contributing parameter. Thus uncertainty propagation, optimization, and design of experiments are straight-forward for complicated models and simulations. \autoref{fig:heat_sobol} provides the dominating parameters necessary for a calculation and the comparison between geometry and constituent material properties. By highlighting interesting and intuitive trends, this plot guides further examinations throughout the remainder of this section. Parameters and resulting properties are grouped by physics such that material dependencies fall on the off-diagonal of the array. In general, geometry dominates over constituent material properties for the majority of composite properties. Because of the unit cell geometry, dependencies on constituents and geometrical parameters are different for the two-directions of anisotropic properties. For example, yarn gap and undulation have alternating strengths between the in-plane and out-of-plane directions and play a large role in the composite anisotropy. Additionally, each composite direction has varying preference for the matrix and fiber phase properties. Standout dependencies highlight intuitive connections: in-plane thermal conductivity and fiber conductivity, out-of-plane permeability and gap, and surface area with respect to the thickness. 

\begin{figure}
	\centering
	\includegraphics[width=0.8\linewidth]{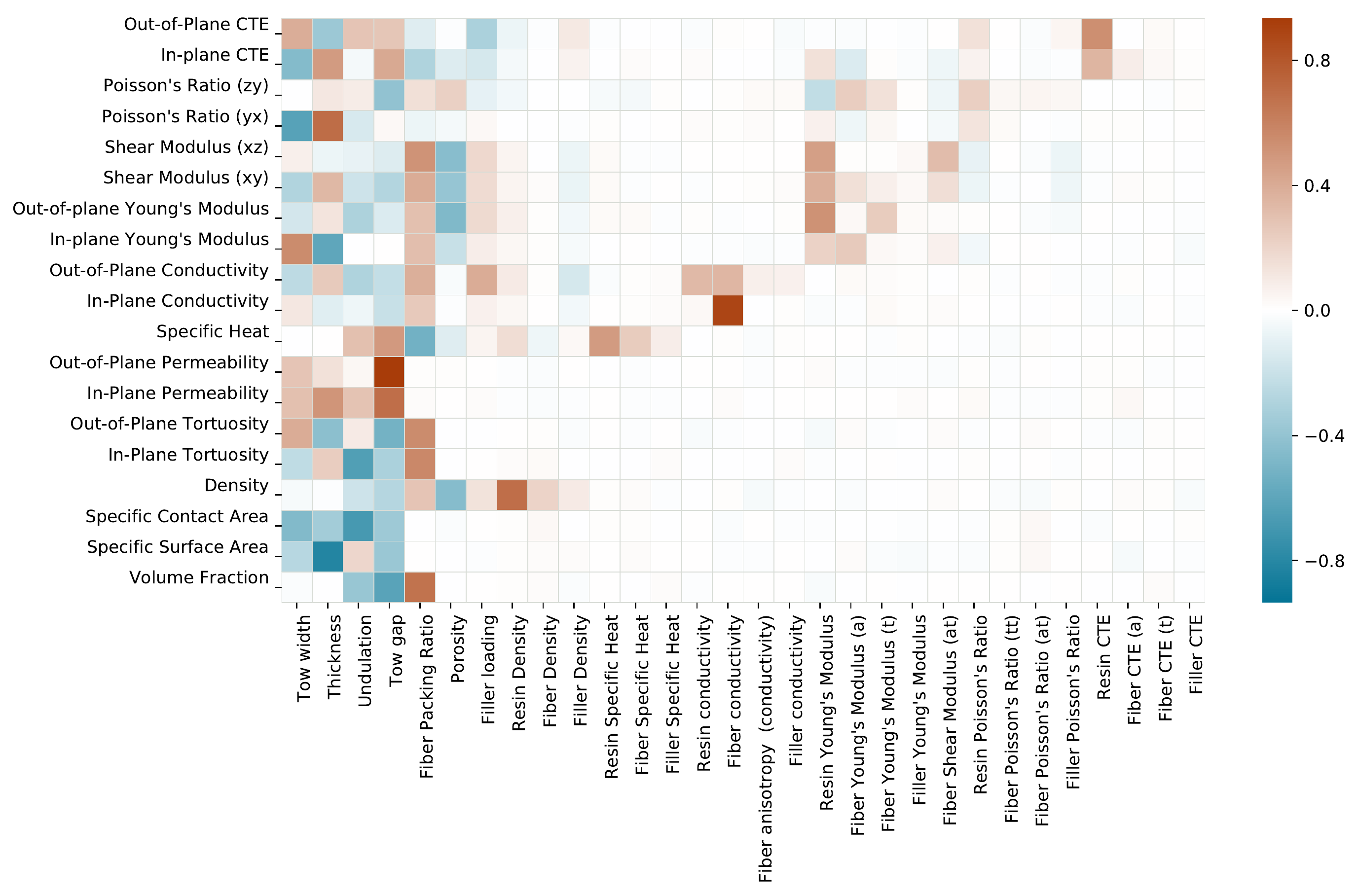}
	\caption{Ranked correlation coefficients arising from the GSA. Note the same general patterns as \autoref{fig:heat_sobol}.}
	\label{fig:heat_corr}
\end{figure}

Correlation coefficients are provided through the LHS study and provide a similar metric for examining the results. \autoref{fig:heat_corr} displays the same qualitative results as Sobol' indices, yet provides signed interactions. Much like Sobol' indices, correlation coefficients display general trends in the results. Along the off-diagonal, positive correlations exist between the composite properties and the associated constituent material properties. However, correlation with geometry is not as consistent, again because of the anisotropic unit cell geometry. Aside from in-plane thermal conductivity and Young's modulus, resin properties dominate over those of the fiber and filler properties. Correlations with geometry also carry though to the anisotropic properties as in-plane and out-of-plane responses largely have opposite dependencies on tow dimensions and undulation value. 

\begin{figure}
	\centering
	\includegraphics[width=0.7\linewidth]{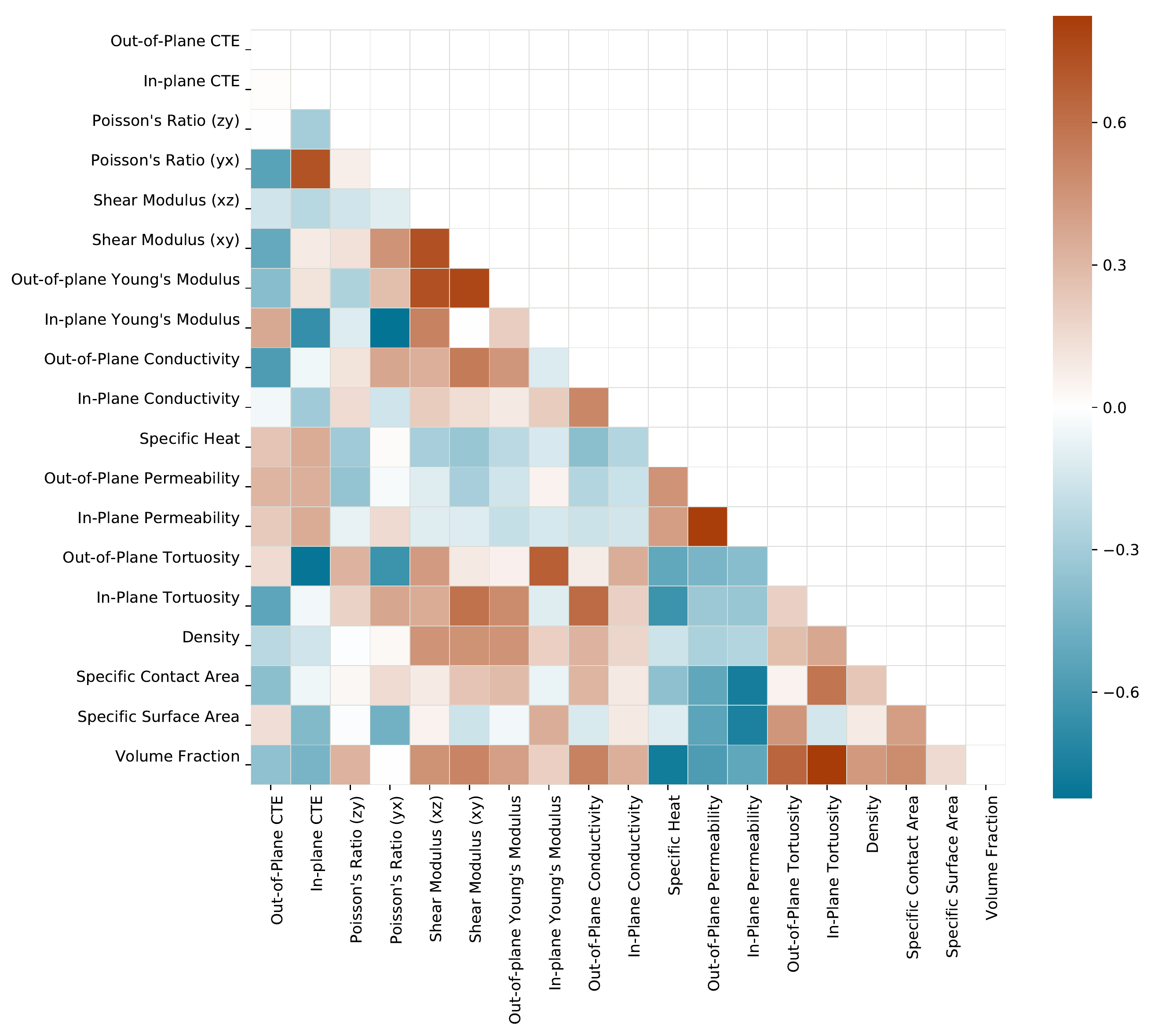}
	\caption{Ranked correlation coefficients between QoIs arising from the GSA.}
	\label{fig:heat_corr_qoi}
\end{figure}

Correlation coefficients between the different QoIs are shown in \autoref{fig:heat_corr_qoi} and present another method for connecting groups of properties. Correlation between properties offers suggestions for optimization and balancing behavior for a given application. Derived geometric properties such as fiber volume fraction indicate which phase is dominant for a QoI. For example, permeability, specific heat, and coefficient of thermal expansion are proportional to matrix phase properties and thus have negative correlations with those favoring the fiber phase, such as conductivity. Using Sobol' indices and correlations as an indicator for trends, we analyze the results and trends for each set of physics in the following subsections.

\subsection*{Fluid flow properties}
\begin{figure}
	\centering 
	\begin{subfigure}{0.4\textwidth}
		\centering
		\includegraphics[width=\linewidth]{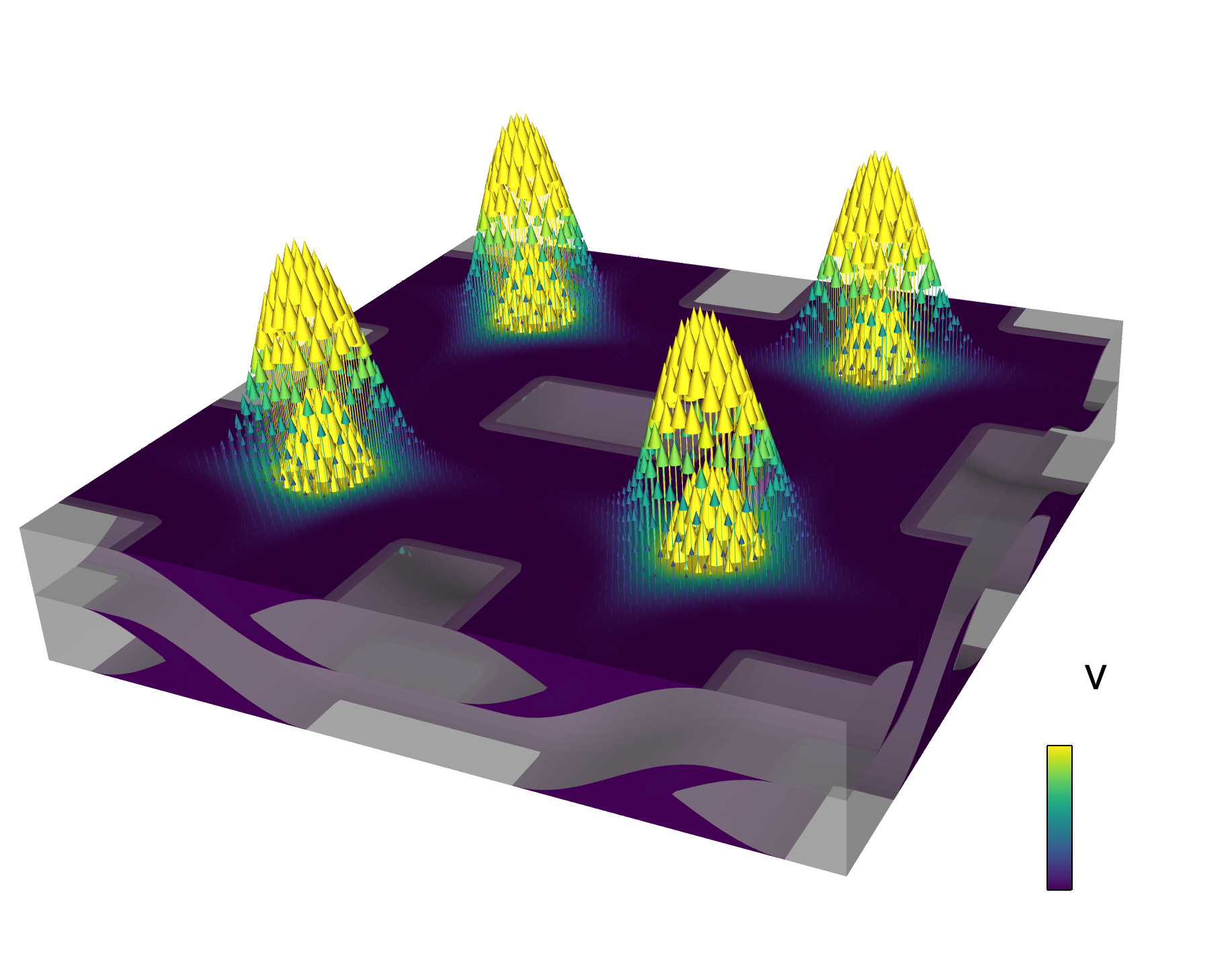}
		\caption{}
		\label{fig:oopperm_example:velocity}
	\end{subfigure}
	\begin{subfigure}{0.4\textwidth}
		\centering
		\includegraphics[width=\linewidth]{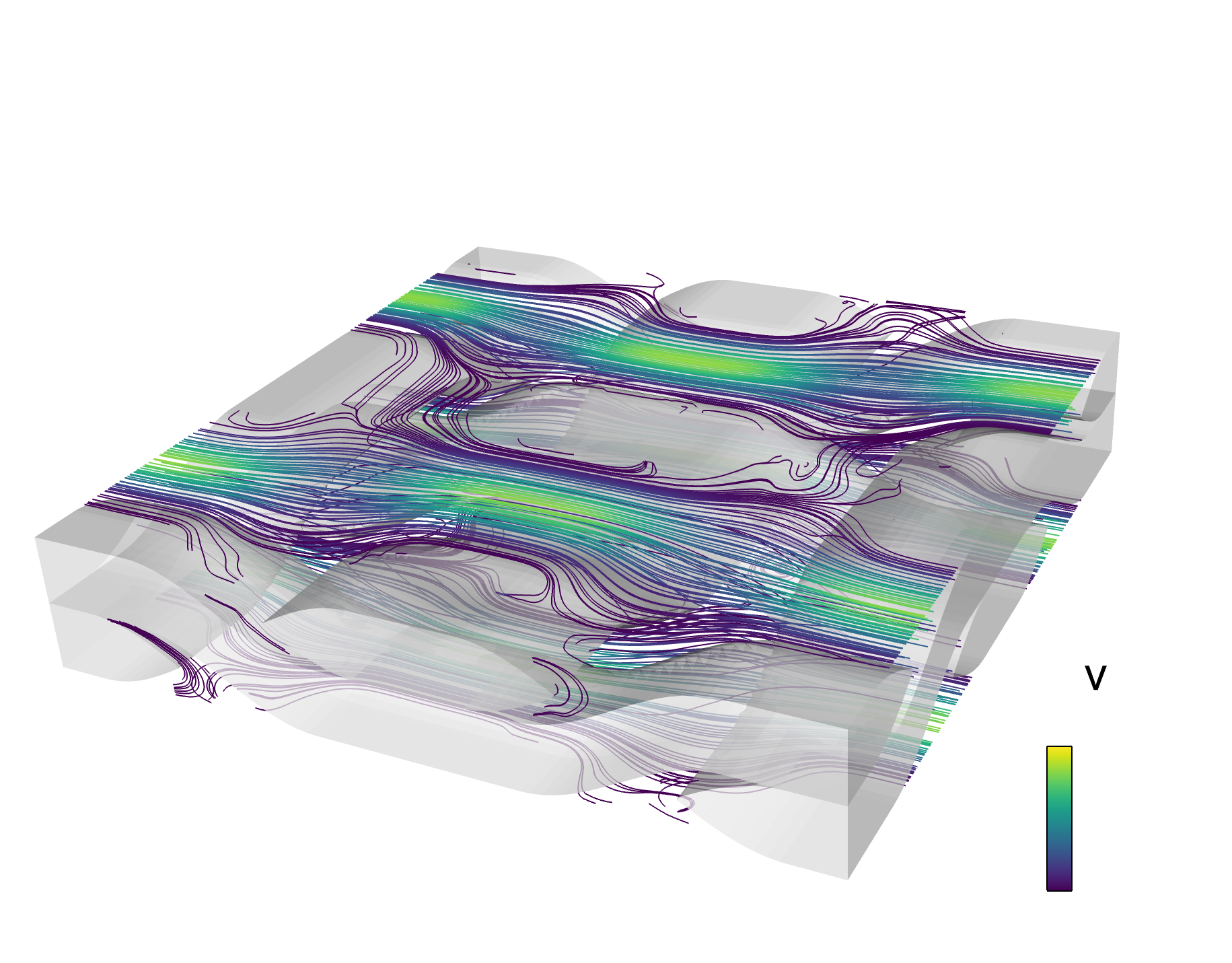}
		\caption{}
		\label{fig:ipperm_example:velocity}
	\end{subfigure}
	\caption{Example velocity field and streamlines for (\subref{fig:oopperm_example:velocity}) out-of-plane and (\subref{fig:ipperm_example:velocity}) in-plane permeability calculations by evaluating fluid flow around the tows.}
	\label{fig:oopperm_example}
\end{figure}

Results for the nominal permeability simulation are visualized in \autoref{fig:oopperm_example}. In the out-of-plane direction, \autoref{fig:oopperm_example:velocity}, the flow is well-formed and uniform through the thickness. Transport is confined to square-shaped gap regions between yarns, with minimal movement in other portions of the domain. Additionally, there is an increase in velocity at the narrowest point between half-layers with a corresponding spike in pressure at the unit cell midplane. In the in in-plane direction, \autoref{fig:ipperm_example:velocity}, flow occurs in the tube-like channels between layers of fabric along the surface of the unit cell. Increased velocity occurs at the constrictions where perpendicular tows cross. 

\begin{figure}
	\centering
	\begin{subfigure}{.32\textwidth}
		\centering
		\includegraphics[width=\linewidth]{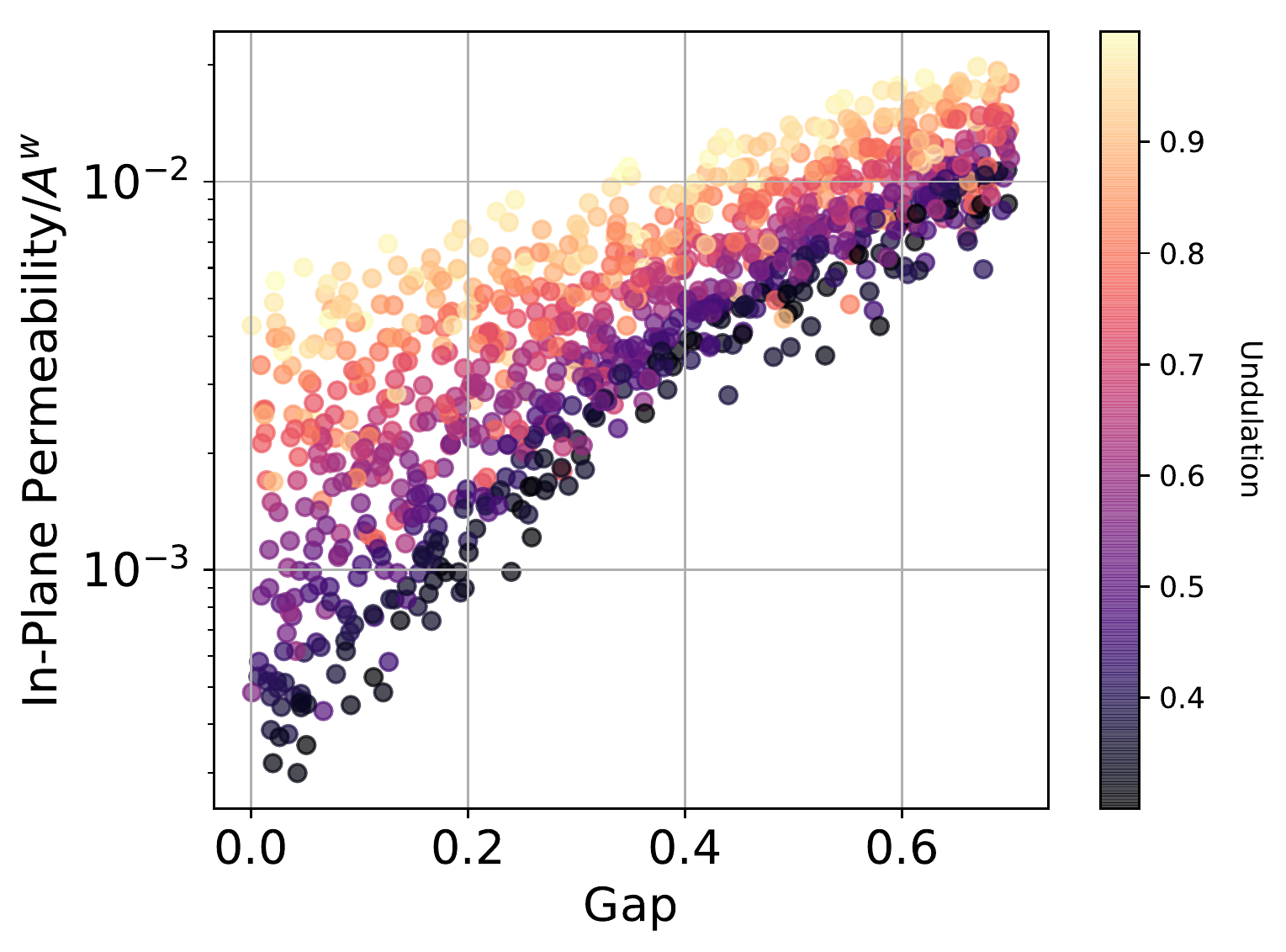}
		\caption{}
		\label{fig:gap_ipperm2}
	\end{subfigure} 
	\begin{subfigure}{.32\textwidth}
		\centering
		\includegraphics[width=\linewidth]{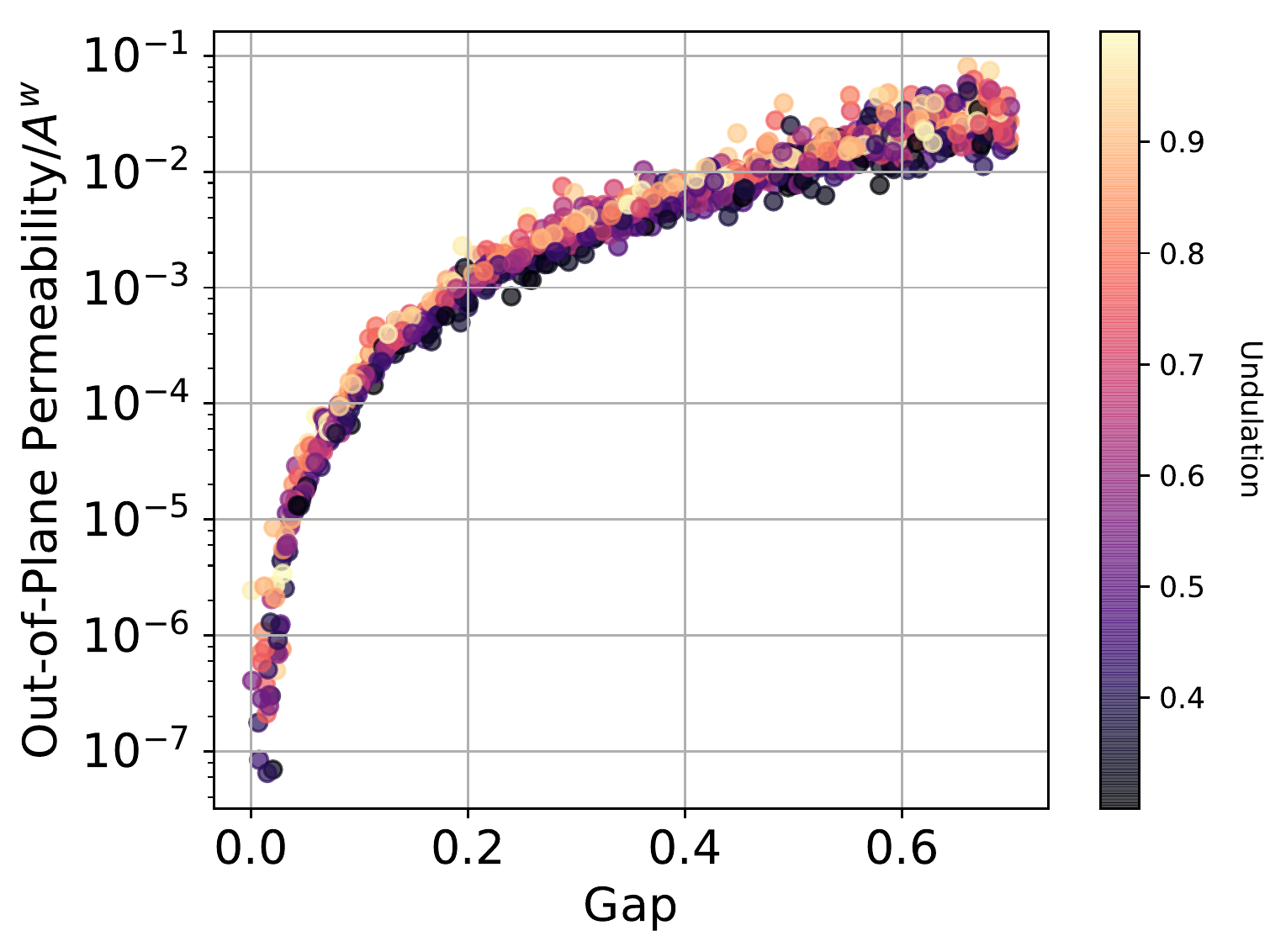}
		\caption{}
		\label{fig:gap_oopperm2}
	\end{subfigure} 
	\begin{subfigure}{.32\textwidth}
		\centering
		\includegraphics[width=\linewidth]{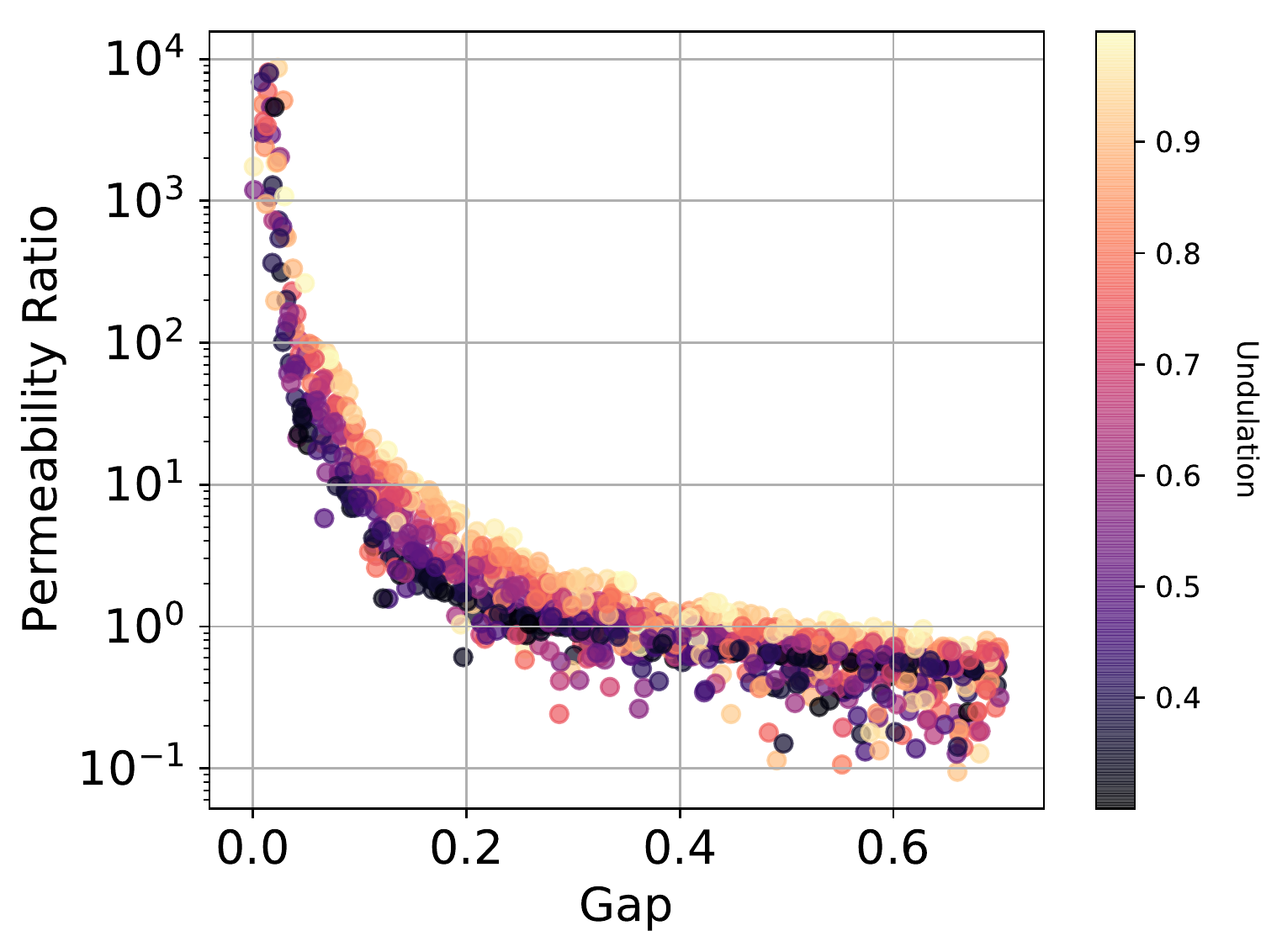}
		\caption{}
		\label{fig:gap_ani}
	\end{subfigure}
	\caption{Summary of weave permeability as a function of the gap between tows. Permeability (\subref{fig:gap_ipperm2}) in-plane, (\subref{fig:gap_oopperm2}) out-of-plane, scaled by area. (\subref{fig:gap_ani}) shows the ratio between the two directions. Data is colored by undulation.}
	\label{fig:perm1}
\end{figure}

Scaling by cross-sectional area in \autoref{fig:perm1} condenses the permeability values by minimizing the scatter from parameters that determine the cross-sectional area $A\phSc{w}{}$ (\autoref{eq:A_cross_section}): yarn thickness, width, and undulation. Undulation determines the amount of a yarn's path spent transitioning over a perpendicular tow, as well as the sinusoidal shapes on the cross-section. Sobol' indices and correlation coefficients indicate a strong dependence of permeability on the yarn gap which is reinforced by the resulting trends. Out-of-plane permeability (\autoref{fig:oopperm_example:velocity}) exponentially vanishes as the gap is decreased, and a secondary dependence on undulation is illustrated with color. In-plane permeability, however, remains finite as the gap goes to zero, as in-plane flow (\autoref{fig:ipperm_example:velocity}) is largely dictated by the undulation parameter, which determines the overall shape and volume of the channels between the yarns. Thus, it is strongly correlated with the scaling between permeability in the two directions. At high undulation values, larger flow channels are available and both in-plane and out-of-plane permeability is increased.

\begin{figure}
	\centering 
	\begin{subfigure}{.4\textwidth}
		\centering
		\includegraphics[width=1.\linewidth]{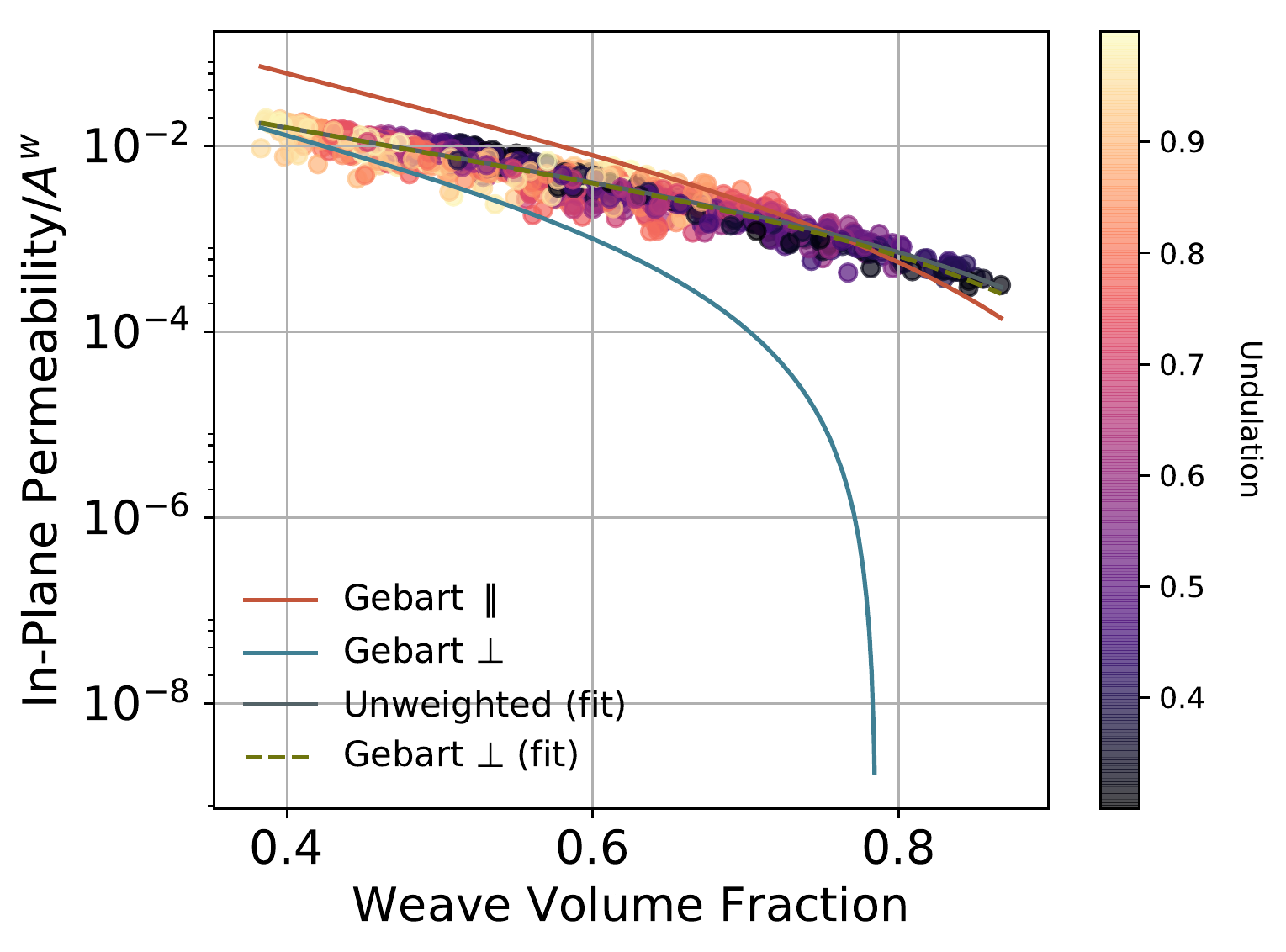}
		\caption{}
		\label{fig:ip_perm_fit}
	\end{subfigure}
	\begin{subfigure}{.4\textwidth}
		\centering
		\includegraphics[width=1.\linewidth]{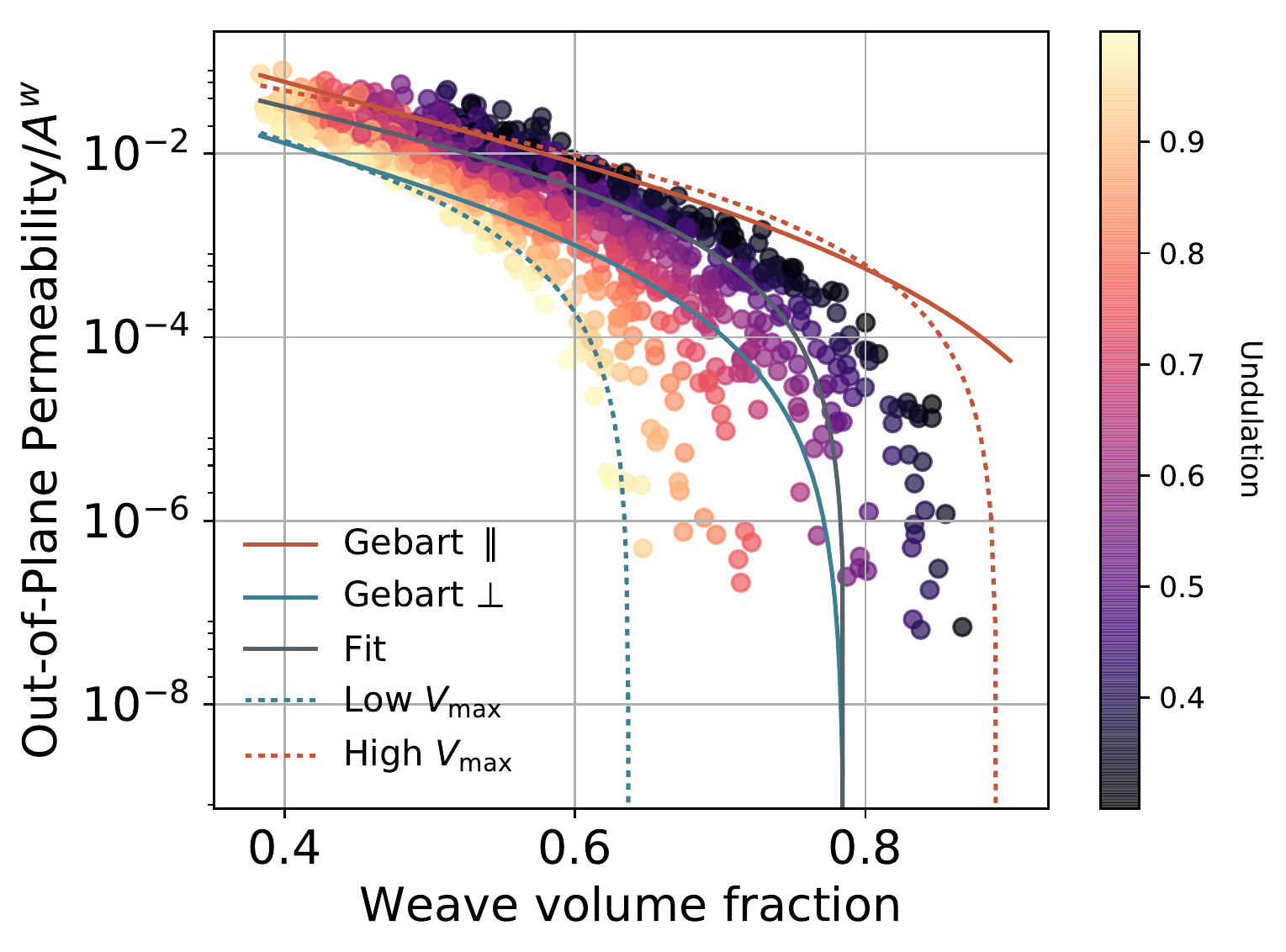}
		\caption{}
		\label{fig:oop_perm_fit}
	\end{subfigure}
	\caption{(\subref{fig:ip_perm_fit}) in-plane and (\subref{fig:ip_perm_fit}) out-of-plane permeability compared to Gebart's approximations for perpendicular (\autoref{eq:gebart_perp}) and parallel (\autoref{eq:gebart_par}) cylinder arrangements for a quadratic array. The remaining comparisons are obtained using variations of (\autoref{eq:gebart_perp}).}
	\label{fig:perm2}
\end{figure}

By treating the yarns as impermeable, we can utilize models describing the permeability of arrays of cylinders and describe the dependence on weave volume fraction $v\phSc{w}{}$ seen in \autoref{fig:perm1}. Gebart \cite{gebart1992} presents results for unidirectional arrays of cylinders in hexagonal and quadratic arrangements in the longitudinal direction:
\begin{equation}
\label{eq:gebart_par}
\frac{\kappa_{\parallel}}{A\phSc{w}{}} = \frac{8}{\pi c}\frac{(1-v\phSc{w}{})^3}{{v\phSc{w}{}}^2},
\end{equation}
and in the transverse direction:
\begin{equation}
\label{eq:gebart_perp}
\frac{\kappa_{\perp}}{A\phSc{w}{}} = \frac{C_1}{\pi}\left(\sqrt{\frac{V_{\max}}{v\phSc{w}{}}}-1\right)^{5/2}.
\end{equation}
Here $v\phSc{w}{}$ represents the weave volume fraction and $A\phSc{w}{}$ represents the nominal cross-sectional area. $V_{\max}$ represents a maximum volume fraction of solid phase, and parameters $C_1$ and $c$ depend on fiber arrangement and volume fraction.

For in-plane permeability, \autoref{fig:ip_perm_fit}, Gebart's models envelope the results, as half of the yarns in a unit cell are parallel to flow and the other half are perpendicular. We fit a weighted average of the two models to the data, using the approach presented by Mattern et al. \cite{mattern2008,jackson1986,woudberg2017}, shown with the gray line in \autoref{fig:ip_perm_fit}. However, when using $C_1$ and $V_{max}$ as fitting parameters, the perpendicular model reasonably captures the mean of the permeability results with $C_1=0.147$ and $V_{max} = 1$, seen through the dashed line in \autoref{fig:ip_perm_fit}. The models (\autoref{eq:gebart_par}) and (\autoref{eq:gebart_perp}) do not match the permeability drop-off at low porosities and have limited applicability here. Finally, the secondary dependency on undulation illustrates that higher volume fractions are possible with lower values of undulation, typically associated with higher fabric crimp or cure pressure.

For out-of-plane permeability, \autoref{fig:oop_perm_fit}, we focus on the comparison to the perpendicular approximation since all of the tows are perpendicular to the flow, aside from the undulating portions. The obtained results follow the drop-off predicted by the perpendicular approximation at low porosity values. The undulation parameter dictates through-thickness orientation of the yarns and therefore controls the spread in permeability data. Fitting \autoref{eq:gebart_perp} to the obtained data yields $C_1=0.63$ and $V_{max} = 0.79$ and does a better job of capturing the observed behavior than using the parameters presented by Gebart for a square array. Alternatively, $V_{max}$ can be approximated using the analytical weave cross-sectional area, \autoref{eq:A_cross_section}, divided by the area of a rectangle bounding the yarn, $tw$:
\begin{align}
V_{\max} = \left(1-u\left(1-\frac{2}{\pi}\right)\right),
\end{align}
i.e., the area density of yarn cross section. Using the limits of the undulation parameter $u$ yields values of $V_{max}=[0.64,0.89]$. When applied to \autoref{eq:gebart_perp}, the approximation captures the spread of observed permeability. 

\begin{figure}
	\centering 
	\includegraphics[width=0.8\linewidth]{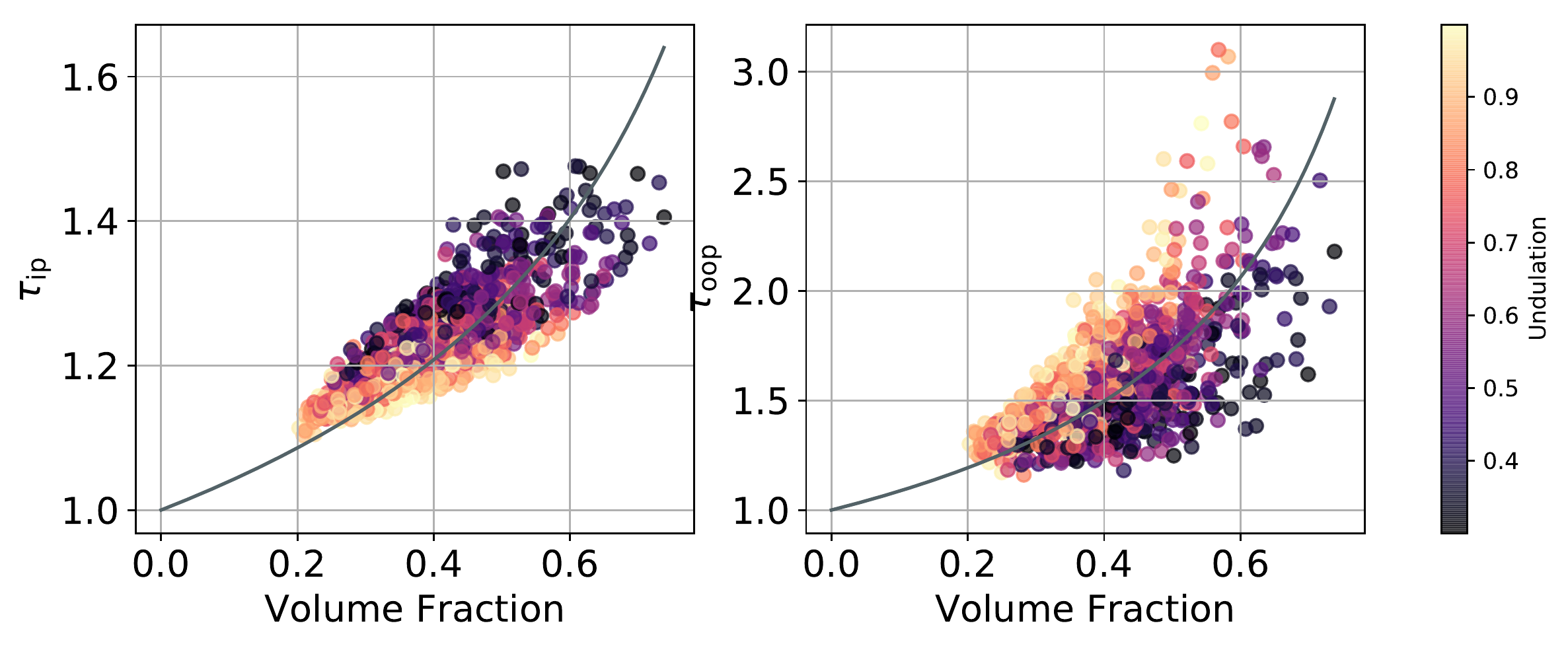}
	\caption{In-plane and out-of-plane tortuosities mixed dependencies on volume fraction and undulation. The Bruggeman approximation, (\autoref{eq:bruggeman}) \cite{bruggeman1935}, has been fit to the data.}
	\label{fig:tort_comp}
\end{figure}

Tortuosity is calculated assuming diffusive transport through the in intra-weave and inter-weave, and thus serves as an alternate approximation to the fluid flow properties of a woven fabric. Tortuosity results are shown as functions of fiber volume fraction and undulation in \autoref{fig:tort_comp}. Intuitively, from the two-scale consideration, tortuosity is dependent on the same geometric parameters as permeability in addition to fiber packing ratio. Additionally, Sobol' indices (\autoref{fig:heat_sobol}) and correlation coefficients (\autoref{fig:heat_corr}) indicate a stronger dependence on undulation over permeability, likely because of the additional transport through yarns. In-plane, higher undulation values result in shorter diffusion path lengths from transport along the porous tows, and lower tortuosities for a given volume fraction. Conversely, for out-of-plane tortuosity, lower undulation values increase the yarn orientation out-of-plane and thus typically have lower tortuosities. 

The Bruggeman approximation \cite{bruggeman1935} describes tortuosity as function of solid volume fraction:
\begin{align}
\tau = (1-\phi)^{(1-\alpha)}.
\label{eq:bruggeman}
\end{align}
Best-fit in-plane and out-of-plane exponents are 1.37 and 1.79, respectively, and are shown with the curves in \autoref{fig:tort_comp}. However, it is impossible to capture the spread and dependencies caused by shape parameter details using traditional predictions on tortuosity. 

\subsection*{Thermal properties}
\begin{figure}
	\centering
	\includegraphics[width=0.4\linewidth]{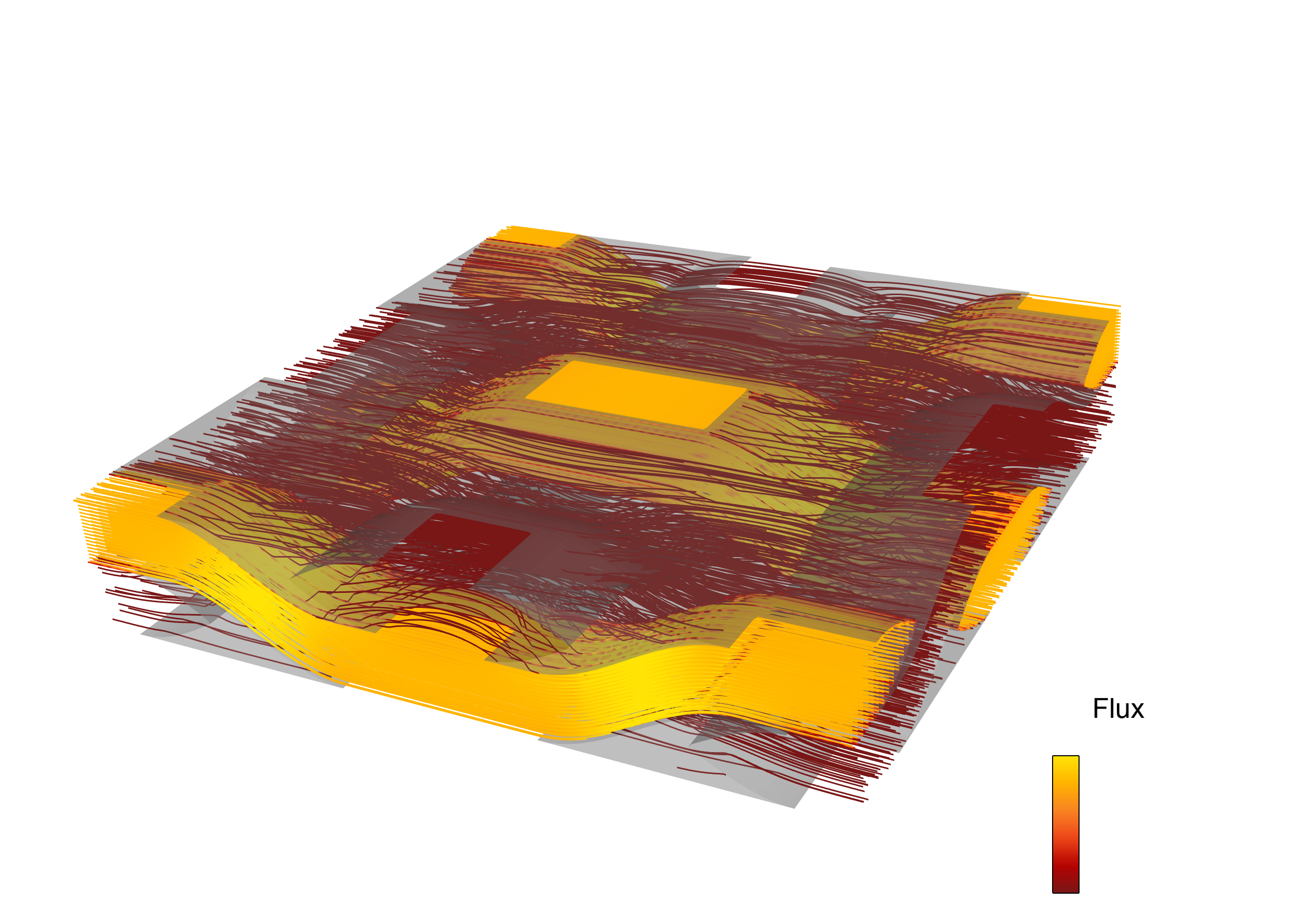}
	\caption{Heat transport through the unit cell is confined to the weave. The resulting heat flux has been visualized through streamlines colored by the flux magnitude.}
	\label{fig:k_ip_example}
\end{figure}

The resulting heat flux from a thermal simulation is visualized through streamlines in \autoref{fig:k_ip_example}. The majority of the heat flux is confined to the yarns aligned with the temperature gradient. The heat flux emphasizes the dependence of effective thermal conductivity on the geometric parameters describing the yarn path and fiber conductivity. The increased flux at the undulating portions of the weave results from maintaining a uniform temperature gradient across the unit cell and the lack of perpendicular yarns that are more conductive than the matrix. Additionally, there is a slightly reduced cross-sectional area in the undulating portions from constructing the tows through translation rather than extrusion.

\begin{figure}
	\centering	
	\includegraphics[width=0.8\linewidth]{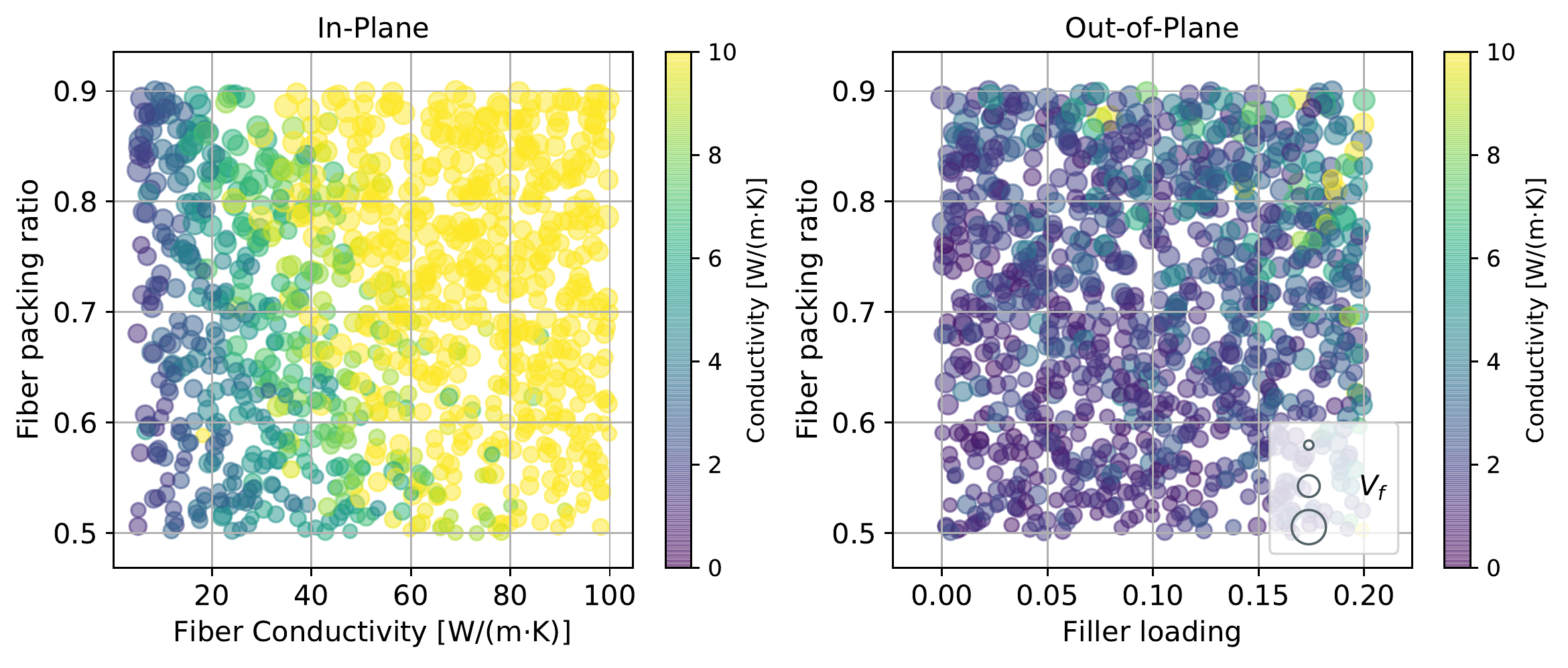}
	\caption{Predominant factors impacting effective thermal conductivities, as identified by Sobol' indices. Marker size is reflective of fiber volume fraction, $v\phSc{*}{f}$, and colored by composite conductivity.}
	\label{fig:cond_scatter}
\end{figure}

Geometric parameters and constituent material properties carry similar magnitude contributions to the composite thermal properties. \autoref{fig:cond_scatter} highlights the most dominant factors and their combined effect on the different conductivity value. In the in-plane direction, from the Sobol' indices (\autoref{fig:heat_sobol}), the dominating factors focus on the fiber phase: fiber conductivity and fiber packing, with the former clearly being the strongest. However, out-of-plane conductivity is dominated by filler loading and fiber packing. Filler loading aids in the conductivity of the matrix phase (more so than resin conductivity) and fiber packing is involved because of the strong contribution from the undulating portions of the weave.

\begin{figure}
	\centering
	\begin{subfigure}{.4\textwidth}
		\centering
		\includegraphics[width=\linewidth]{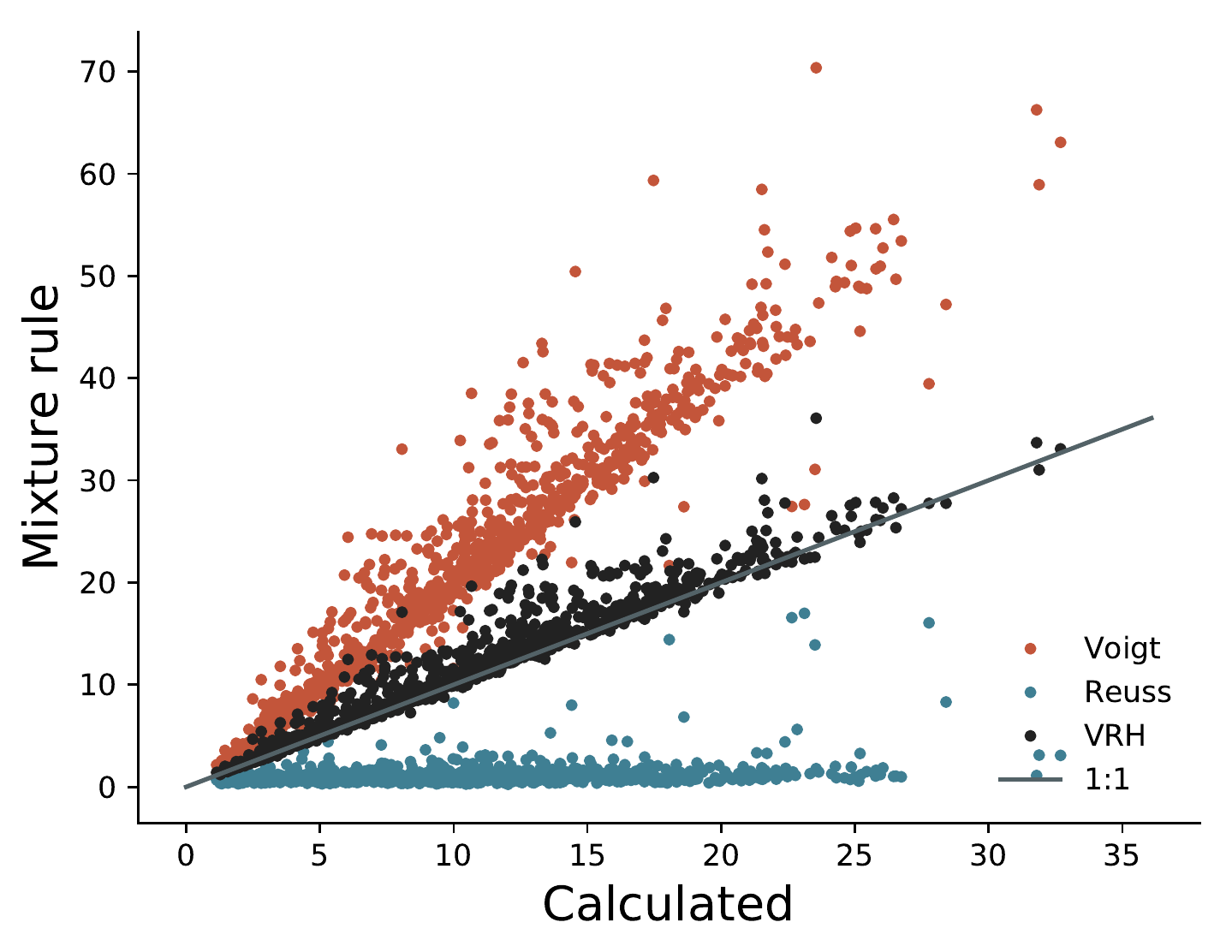}
		\title{In-plane}
		\label{fig:mix_k_ip}
	\end{subfigure}
	\begin{subfigure}{.4\textwidth}
		\centering
		\includegraphics[width=\linewidth]{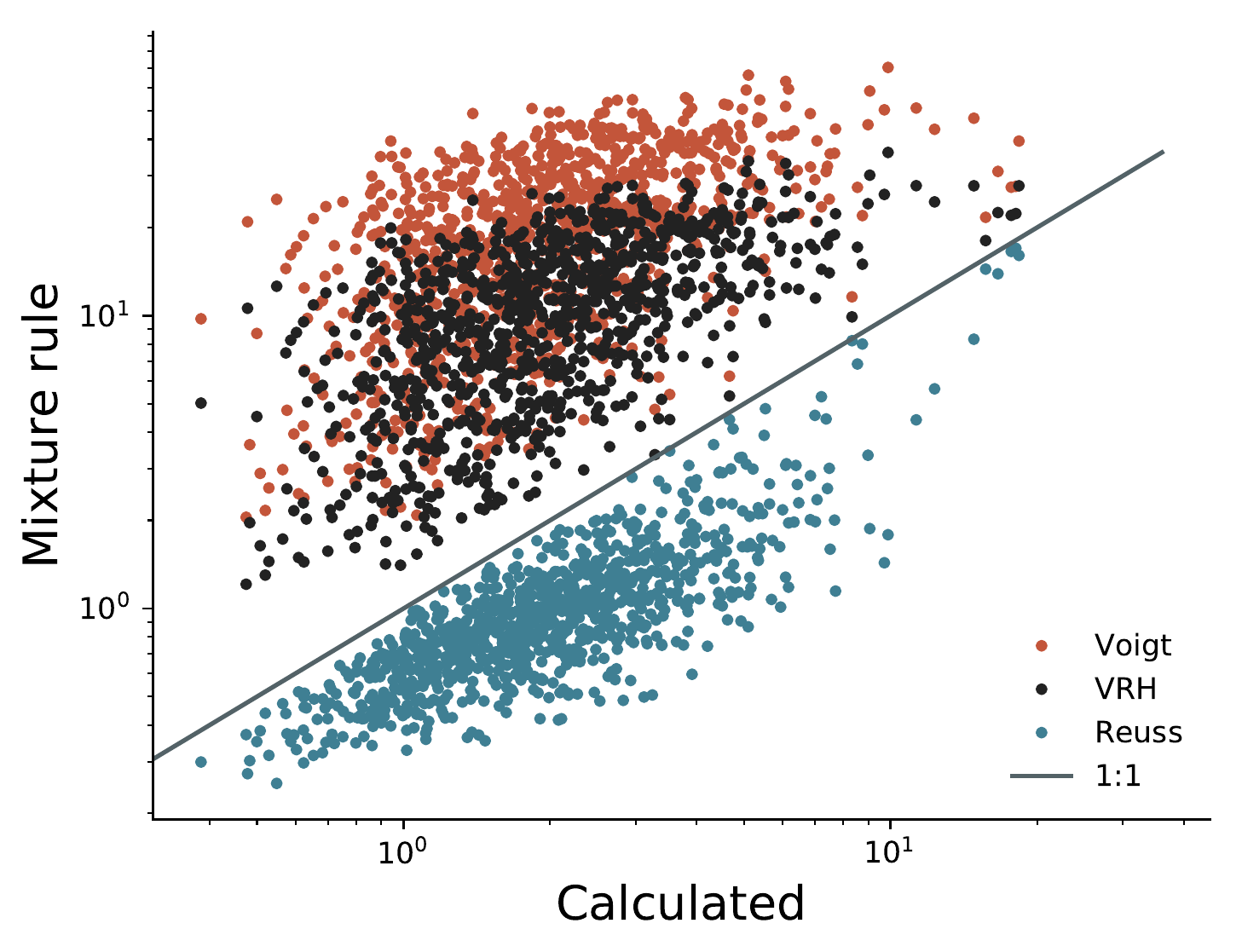}
		\title{Out-of-plane}
		\label{fig:mix_k_oop}
	\end{subfigure}
	\caption{Comparison between calculated composite thermal conductivities and associated Voigt and Reuss bounds based on the mesoscale volume fractions and properties.}
	\label{fig:mixture_rules}
\end{figure}

Voigt ($k\phSc{*}{V}$) and Reuss ($k\phSc{*}{R}$) bounds are common approximations and bounds applied to the behavior of composites assuming isotropic constituents. Bounds are calculated for the composite conductivity by volume averaging the effective matrix properties (calculated from resin, filler, and void) and the effective yarn conductivities in the axial and transverse directions. The Voigt bound (or rule of mixtures) acts as an upper bound on composite properties and models the behavior of composites with inclusions oriented in the direction of loading:
\begin{align}
\label{eq:voigt}
k\phSc{*}{V} &= k\phSc{w}{a} v\phSc{w}{} + k\phSc{m}{} (1-v\phSc{w}{}).
\end{align}
The Reuss bound (inverse rule of mixtures) acts as a lower bound and approximates the behavior of composites with inclusions perpendicular to the direction of loading:
\begin{align}
\label{eq:reuss}
k\phSc{*}{R} &= \left(\frac{v\phSc{w}{}}{k\phSc{w}{t}} +\frac{1-v\phSc{w}{}}{k\phSc{m}{}}\right)^{-1}.
\end{align}
The Voigt-Reuss-Hill (VRH) approximation is an average of the two, arising from various micromechanics communities \cite{chung1963}. In \autoref{fig:mixture_rules}, bound values calculated using each set of trial parameters are plotted against the value obtained from their corresponding effective property simulation. 

For in-plane conductivity, the Voigt and Reuss bounds are satisfied despite the anisotropic constituents: the Voigt values are always greater than computed properties (i.e. the fall above 1:1 line) whereas the Reuss values fall below. Additionally, the VRH approximation proves to be a good estimate of in-plane behavior as half of the weave is inline with the thermal gradient load and half is perpendicular. For through-plane behavior, the bounds are still satisfied yet the VRH approximation doesn't match the data as well.

\subsection*{Elastic properties}
\begin{figure}
	\centering
	\includegraphics[width=0.4\linewidth, draft=false]{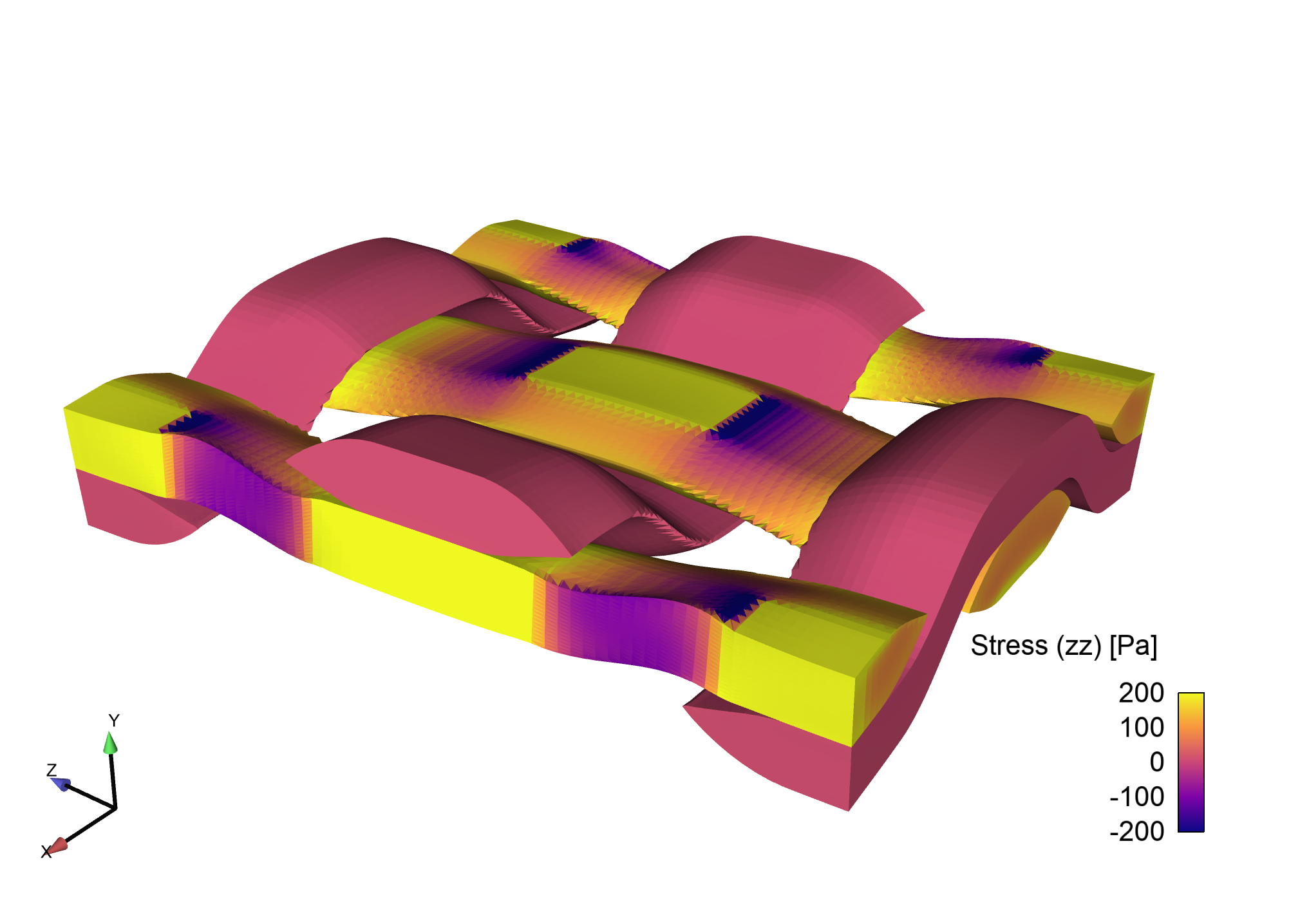}
	\caption{Example resulting stress field from an applied in-plane load.}
	\label{fig:mech_examples}
\end{figure}

Much like thermal conductivity, in-plane mechanical behavior of the composite is dominated by the fibers. An example in-plane Young's modulus calculation is shown in \autoref{fig:mech_examples}, where a unit cell is subjected to an axial load in the $z$-direction. Visualizing the resulting $zz$-stress component in the fibers shows that tows parallel to the load are a major source of stiffness. There are two sources of the compressive stress observed. Bending of the undulating portions of yarns aligned with the load results in compression along the surface. As a result, these portions experience more $z$-displacement than what is prescribed at the entire boundary of the unit cell, thus resulting in the compression seen entirely through the tow thickness at the boundary. Finally, the compliant matrix minimizes stress on perpendicular yarns, which experience the Poisson effect through further bending.

\begin{figure}
	\centering
	\includegraphics[width=0.8\linewidth, draft=false]{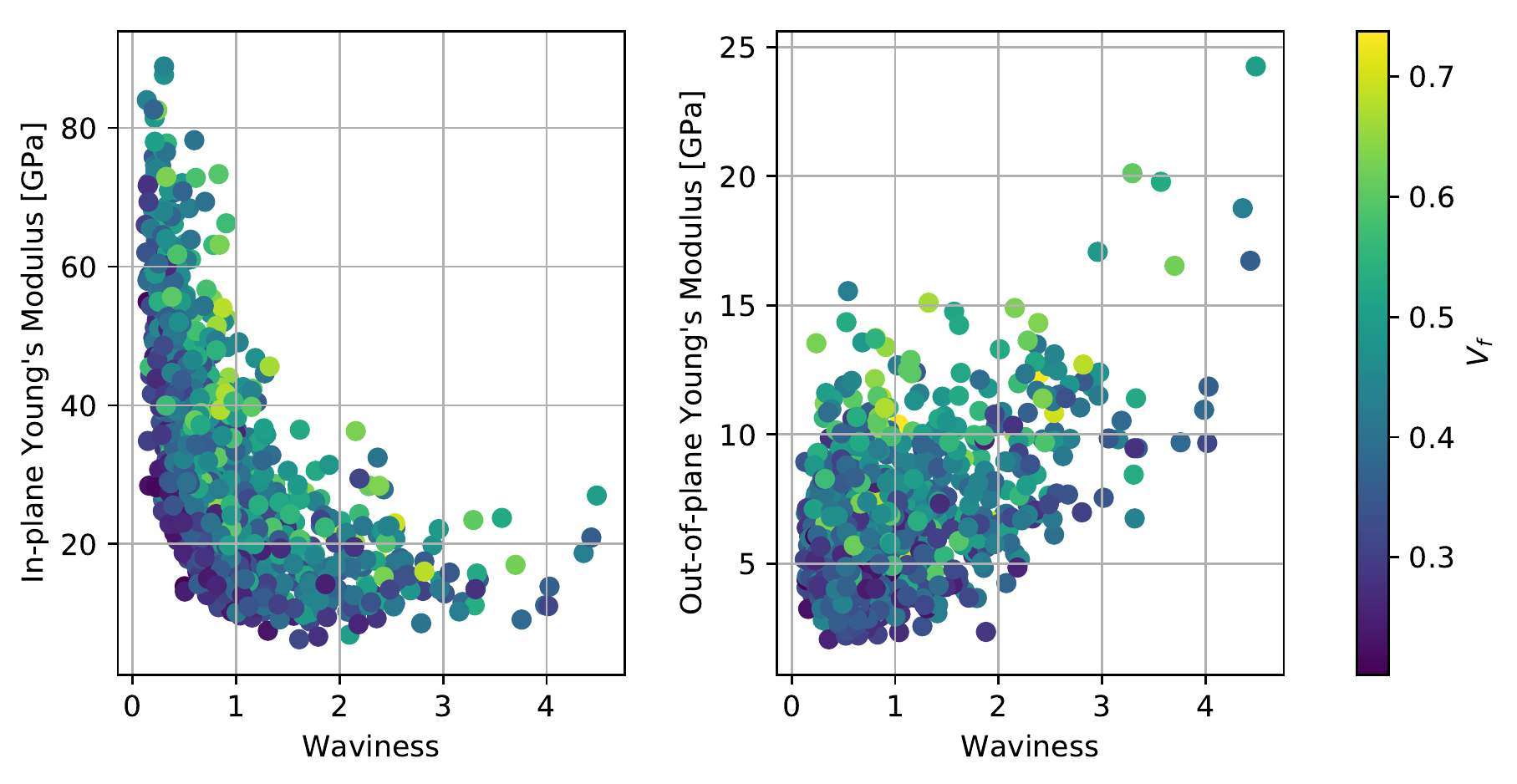}
	\caption{Summary of elastic moduli as a function of waviness. Waviness combines dependencies on different parameters through \autoref{eq:waviness}.}
	\label{fig:mech_geo}
\end{figure}

The impact of the undulating portions of the tows is captured in \autoref{fig:mech_geo}, where each of the Young's moduli is plotted against waviness, \autoref{eq:waviness}, and fiber volume fraction. For the in-plane simulation, bending in the undulating portion is facilitated by the tow waviness. Thus, the in-plane stiffness sharply decreases with increasing waviness, although this effect is less pronounced at higher fiber volume fractions. Conversely, as waviness increases, more fibers are aligned perpendicular to the plane, and out-of-plane stiffness increases from the fiber's axial contribution. Dependence on fiber volume fraction is additionally evidenced through correlation coefficients. 

\begin{figure}
	\centering
	\begin{subfigure}{.4\textwidth}
		\centering
		\includegraphics[width=\linewidth]{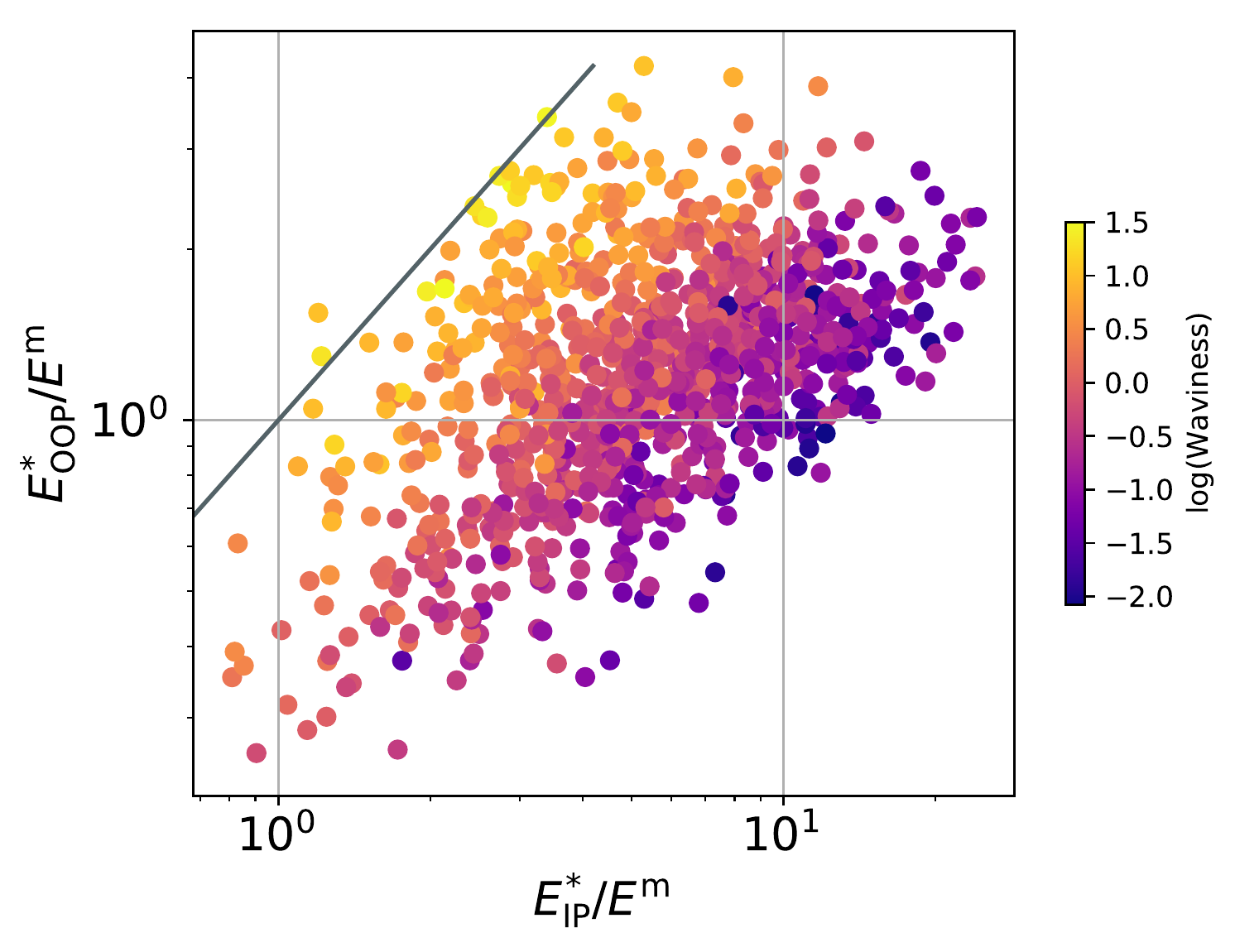}
		\caption{}
		\label{fig:E_ani}
	\end{subfigure}
	\begin{subfigure}{.4\textwidth}
		\centering
		\includegraphics[width=\linewidth]{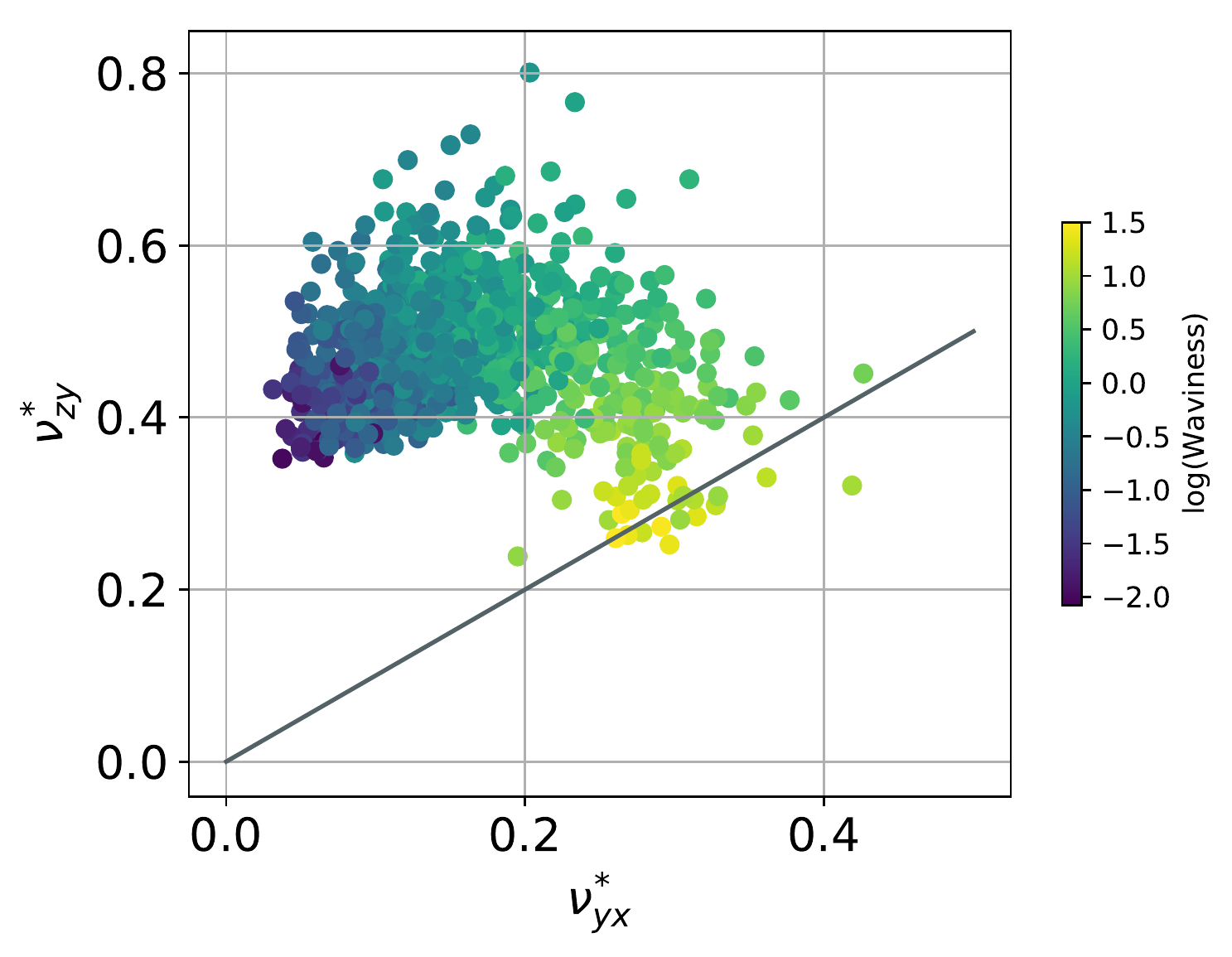}
		\caption{}
		\label{fig:nu_ani}
	\end{subfigure}%
	\caption{Mechanical anisotropy of the composite. (\subref{fig:E_ani}) Young's moduli scaled by the effective matrix stiffness and (\subref{fig:nu_ani}) Poisson's ratio colored by waviness. Solid lines have a slope of one.}
	\label{fig:mech_ani}
\end{figure}

Mechanical isotropy and the impact of waviness is illustrated in \autoref{fig:mech_ani}. Normalizing by the matrix modulus isolates dependence on the tows and fibers. WWe find that the in-plane stiffness is always higher than the out-of-plane stiffness, an intuitive result, and that composite isotropy increases with waviness. Poisson's ratios exhibit a similar dependence on waviness. 

\begin{figure}
	\centering
	\includegraphics[width=0.4\linewidth]{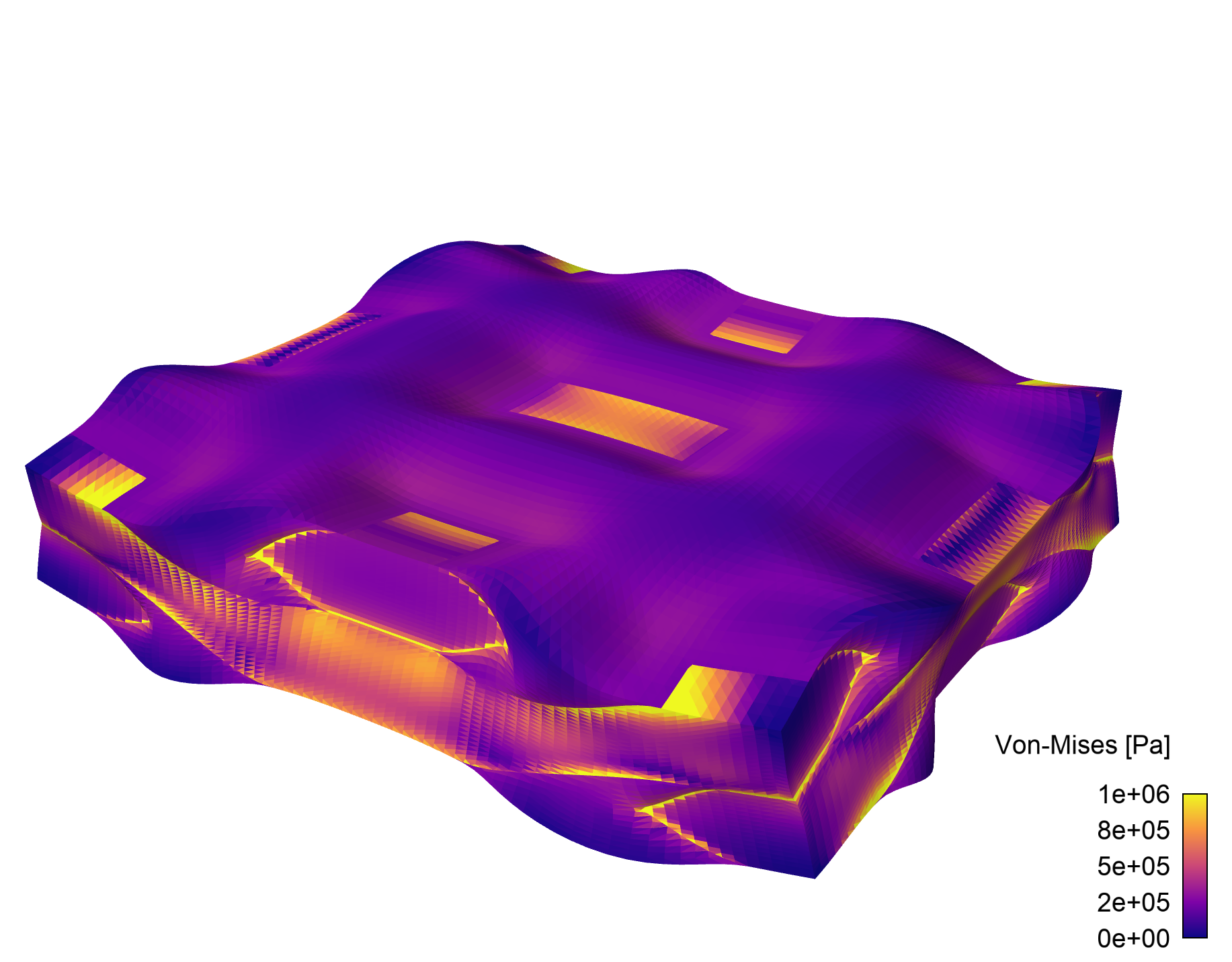}
	\caption{Von Mises stress distribution resulting from thermal expansion.}
	\label{fig:cte_example}
\end{figure}

Under the large thermal gradients experienced by a TPS, expansion becomes a critical role in its performance and longevity. The expansion of an example unit cell under a thermal load is presented in \autoref{fig:cte_example}, where expansion is dominated by the matrix phase. Correspondingly, the weave straightens out to accommodate the overall expansion of the unit cell. According to Sobol' indices (\autoref{fig:heat_sobol}), thermal expansivity is clearly dominated by the resin CTE and has a negative correlation with fiber volume fraction due to the fiber's contraction hindering the matrix's swelling. Additionally, yarns show higher residual stress due to their minimal expansion and high stiffness. Correlations with elastic moduli indicate that in-plane expansivity is hindered by in-plane stiffness, yet increases with higher out-of-plane modulus. These correlations are reversed for out-of-plane expansivity. 

\begin{figure}
	\centering
	\includegraphics[width=0.4\linewidth]{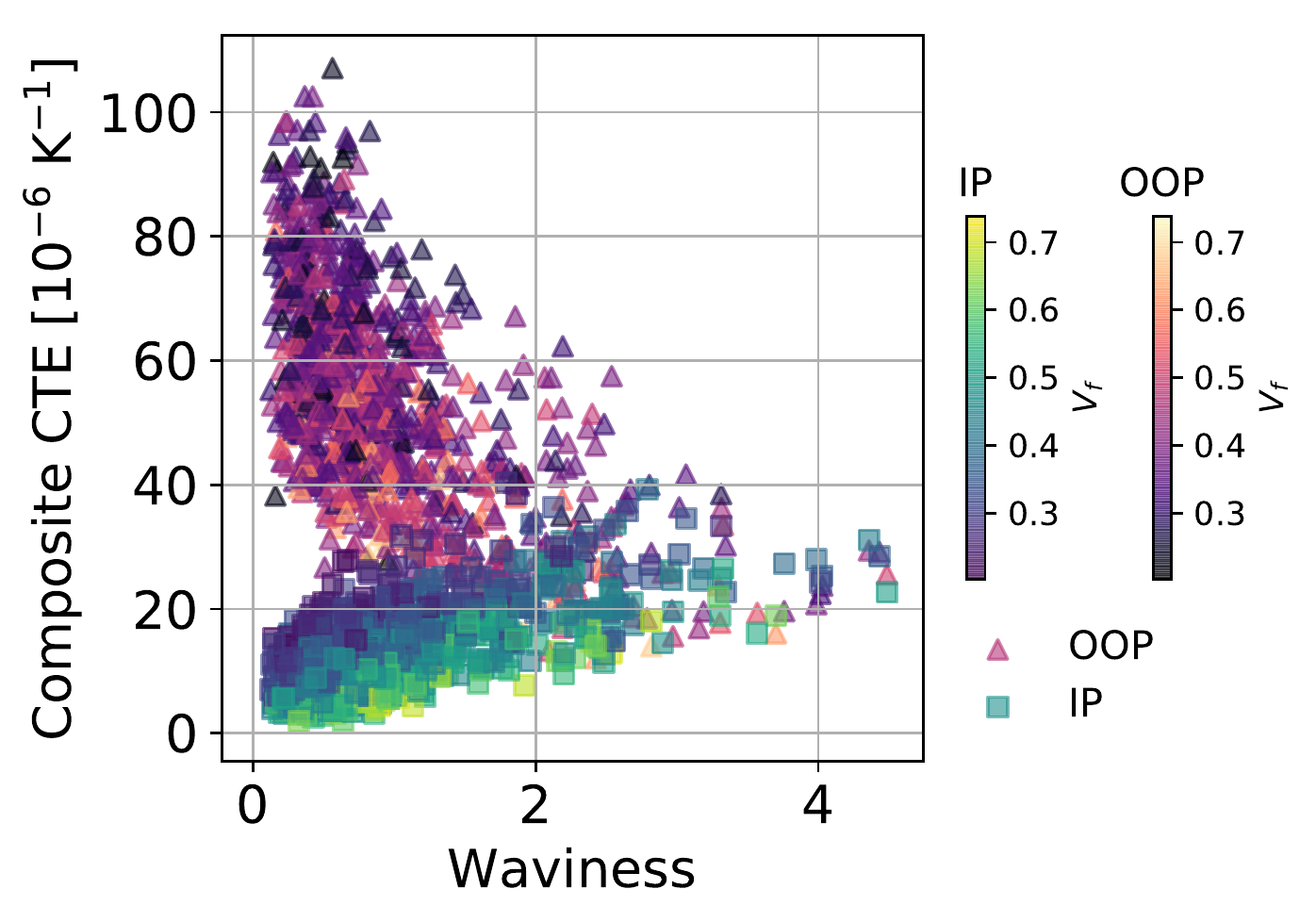}
	\caption{CTE versus waviness, colored by volume fraction.}
	\label{fig:CTE_wavi}
\end{figure}

The dependence of expansivity on waviness and fiber volume fraction is shown in \autoref{fig:CTE_wavi}. As expected, isotropy increases with waviness, yet at low waviness values the out-of-plane expansivity is typically much larger. In both directions, increased fiber volume fraction decreases both the expansivity and the dependency on waviness.

\section{Conclusions}
This work presents a methodology for characterizing the design space of woven composites across different physical properties. An approach using simulations at the mesoscale connects constituent material properties at the microscale to composite effective behavior at the macroscale. By fully characterizing this relation between design space and effective properties, material design can be optimized for a given application. 

Uniform ranges of geometric and material properties were sampled to inform finite element simulations of a mesoscale unit cell consisting of the fabric weave and surrounding matrix for determining relevant effective properties. This included geometric consequences (fiber volume fraction, density, and specific heat), thermal properties (conductivity), flow properties (tortuosity and weave permeability), and mechanical properties (elastic moduli, Poisson's ratio, and thermal expansivity).

The results of this study present both the spread in effective properties from the input ranges, but also interactions and dependencies not available through single property analyses. Sobol' indices obtained through a PCE of the results indicate dominating parameters for a given QoI, and drive the subsequent assessment of the large spread in results. Across the physical properties examined, there are nearly equal dependencies on both the geometry at the mesoscale and the choice of constituent materials. 

As with any numerical simulation, the approach presented here carries certain limitations. Simulation at the mesoscale requires assumptions on the geometry and approximations of the finer scale details of the composite. However, the chosen geometries capture relevant characteristics of the composite to couple geometry and physical properties at a first-order level. The complexity of physical models follows the same rationale --- although higher fidelity simulations are possible, we forgo expensive simulations and modeling to focus on the variability and correlations in effective behavior through a large number of samples. 

This framework offers a useful starting point for future studies in composite analysis. The approach outlined can be generalized and applied to other materials and applications reliant on material behavior at the mesoscale. For thermal protection system applications, this workflow offers a valuable tool for composite material design that forgoes expensive testing and development. 

With stochastic manufacturing conditions, the distribution of yarns and layers of fabric creates nonideal unit cells in the periodic structure, and in turn creates local non-uniformities in physical properties \cite{vanaerschot2017, badel2008, semeraro2020}. Using image-based material geometries offers a source of validation for the present approach and can inform more refined studies in the future.

\section*{Data Availability}

The raw data required to reproduce these findings are available to download from https://doi.org/10.17632/2ng65hbxtj.1. The processed data required to reproduce these findings are available to download from https://doi.org/10.17632/2ng65hbxtj.1.

\section*{Acknowledgements}
We appreciate many helpful conversations with numerous colleagues at Sandia National Laboratories.  In particular, we thank the useful peer review and comments from Leah Tuttle, Jeff Engerer, and Mart\'{i}n Di Stefano.  This work was funded by the US Department of Energy's National Nuclear Security Administration, who did not influence the design, execution, or publication of this manuscript.

Sandia National Laboratories is a multi-mission laboratory managed and operated by National Technology and Engineering Solutions of Sandia, LLC., a wholly owned subsidiary of Honeywell International, Inc., for the U.S. Department of Energy’s National Nuclear Security Administration under contract DE-NA0003525. This paper describes objective technical results and analysis. Any subjective views or opinions that might be expressed in the paper do not necessarily represent the views of the U.S. Department of Energy or the United States Government.

\normalfont 
\bibliographystyle{elsarticle-harv}
\bibliography{references}

\end{document}